\RequirePackage{ifpdf}
\documentclass[12pt,letterpaper]{article}
\pdfoutput=1
\usepackage{jheppub}
\usepackage{epsfig}
\usepackage{enumitem}
\usepackage[latin1]{inputenc}
\usepackage{bbm,amsfonts}
\usepackage{rotating,graphicx}
\usepackage{amssymb,amsmath,amsfonts,mathtools}
\usepackage{fancybox,diagbox}
\usepackage{enumerate}
\usepackage{dsfont}
\usepackage{verbatim}
\usepackage{wrapfig}
\usepackage{slashed}
\usepackage{shuffle}
\usepackage{subcaption}
\usepackage{braket}
\usepackage{pdflscape}
\usepackage{accents}
\usepackage{afterpage}

\author{Marco S. Bianchi}
 
\affiliation{Instituto de Ciencias F\'isicas y Matem\'aticas, Universidad Austral de Chile, Casilla 567, Valdivia, Chile}

\emailAdd{marco.bianchi@uach.cl}  
  
\abstract{I consider three-point functions of twist-one operators in ABJM at weak coupling. 
I compute the structure constant of correlators involving one twist-one un-protected operator and two protected ones for a few finite values of the spin, up to two-loop order.
As an application I enforce a limit on the gauge group ranks, in which I relate the structure constant for three chiral primary operators to the expectation value of a supersymmetric Wilson loop. Such a relation is then used to perform a successful five-loop test on the matrix model conjectured to describe the supersymmetric Wilson loop. 
}

\title{On three-point functions in ABJM and the latitude Wilson loop} 

\newcommand{\coleq}{\underset{\begin{minipage}{0.65cm}\centering \scriptsize color\\[-3pt] limit\end{minipage}}{=}}

\newcommand{\be}{\begin{equation}}
\newcommand{\ee}{\end{equation}}
\newcommand{\beq}{\begin{equation}}
\newcommand{\eeq}{\end{equation}}
\newcommand{\bea}{\begin{eqnarray}}
\newcommand{\eea}{\end{eqnarray}}
\newcommand{\ena}{\end{eqnarray}}

\def\Tr{\textrm{Tr}}

\numberwithin{equation}{section}

\def\clock{{\count0=\time
           \divide\count0 60
           \ifnum\count0<10 0\fi\the\count0
           \multiply\count0 -60 \advance\count0 \time
           :\ifnum\count0<10 0\fi \the\count0
         }}
\newcommand{\timestamp}{{\small\vbox{\hbox{\tt\jobname.tex}
\hbox{\the\day/\the\month/\the\year, \clock}}}}
\newlength{\dhatheight}
\newcommand{\doublehat}[1]{%
    \settoheight{\dhatheight}{\ensuremath{\hat{#1}}}%
    \addtolength{\dhatheight}{-0.35ex}%
    \hat{\vphantom{\rule{1pt}{\dhatheight}}%
    \smash{\hat{#1}}}}

\begin{document}

\maketitle
\allowdisplaybreaks

\section{Introduction and summary}

In this note I approach the problem of computing quantum corrections to three-point functions of local twist-one operators in the ABJM theory \cite{Aharony:2008ug} at weak coupling.
Three-point functions are central objects in conformal field theories. Their determination is notoriously hard from direct computation.

These difficulties can be sidestepped in certain fortunate cases, for example retrieving information on them from the OPE decomposition of higher-point correlators \cite{Dolan:2004iy} (usually of protected operators), which can sometimes be easier to determine.
In the special case of ${\cal N}=4$ SYM theory in four dimensions, a novel approach consists in exploting the conjectured integrability of the theory for computing its three-point functions \cite{Basso:2015zoa}.

Since ABJM seems to share integrability properties of ${\cal N}=4$ SYM, at least for the planar spectral problem \cite{Gaiotto:2008cg,Minahan:2008hf,Gromov:2008qe}, there is hope that the integrability framework for three-point functions could be eventually extended to ABJM.
At a difference with ${\cal N}=4$ SYM, ABJM lacks any perturbative data on three-point functions at quantum level whatsoever, except for the computation of the two-loop structure constant of three protected operators of \cite{Young:2014lka,Young:2014sia}. In ABJM these are non-trivial functions of the coupling constant (the first quantum corrections starting at two-loop order), again in contrast to the situation of maximally supersymmetric Yang-Mills in four dimensions, where the corresponding correlators are protected against quantum corrections \cite{Lee:1998bxa,Eden:1999gh,Arutyunov:2001qw}.

Such a scarcity of perturbative information on quantum three-point functions in ABJM is in itself a motivation for their study.
Still, my original reason for addressing this topic stems from a seemingly unrelated subject, that is supersymmetric Wilson loops in the ABJM model.
In particular, I consider  the so-called latitude Wilson loop in ABJM \cite{Bianchi:2014laa} and a conjecture on its exact expectation value via a matrix model descritpion \cite{Bianchi:2018bke}.
While evidence has been collected in favor of such a proposal \cite{Bianchi:2014laa}, no proof thereof is available, calling for further tests of the conjecture.

Given the difficulty of performing high precision perturbative computation of these Wilson loop expectation value, I aim here to perform an indirect check as follows.
My idea consists in relating a limit of the structure constant for a particular three-point function in ABJM to the matrix model computing the expectation value of the Wilson loop.
Such a relation then serves as a mutual benchmark for the perturbative calculation of the full structure constant at two loops (the first non-trivial quantum correction) and of the matrix model proposal.

The idea behind this connection lies in the existence of a class of diagrams in the perturbative expansion of the supersymmetric Wilson loop,
which effectively contains a three-point function correlator of local length-two chiral primary operators constructed as an R-symmetry traceless combination of the scalar fields. Such graphs originate from the particular structure of the supersymmetric connection of the Wilson loop, which features a coupling to a pair of scalar fields. The details of such a construction are reviewed in section \ref{sec:latitude}.

In order to turn this intuition into a precise statement, I consider a selective multicolor limit on the $U(N_1)\times U(N_2)$ ABJM model \cite{Aharony:2008gk}. In such a limit, where $N_2\gg N_1$, only the planar leading power $N_2^l$ is retained in the quantum correction at $l$ loops for the structure constant, measured after taking the ratio by its tree-level value. 
In such a limit I explicitly relate the expectation value of the latitude Wilson loop at five loops, which can be computed using its alleged description in terms of a matrix model, to the two-loop correction of the given three-point function. This allows to extract a prediction for its two-loop correction (a color component of it, actually) \eqref{eq:prediction}.

I devote the second part of the article to test such a prediction via an explicit perturbative computation of the structure constant at the lowest non-trivial order at weak coupling $k\gg 1$, that is two-loops.
I deploy a method consisting in integrating the three-point function over the insertion point of one of the operators \cite{Plefka:2012rd}.
The punchline is that after applying such an operation on the general structure of the correlator determined by conformal invariance and on the perturbative expansion in terms of Feynman diagrams, the structure constant can be extracted by comparison.
The advantage stems from a drastic simplification that the additional integration (which corresponds to a soft limit in momentum space) triggers at the level of the Feynman diagrams, turning them into two-point function graphs, which are much easier to evaluate. I provide more details on this procedure in section \ref{sec:method}.

In fact, such a line of reasoning had already been adopted in literature for tackling the very same three-point function computation that I am after \cite{Young:2014lka}.
In section \ref{sec:method} I first raise some criticism on such a previous results and explain why I think the method which it was derived with and that I outlined above, can be plagued, in some pathological cases such as the present problem, by some fallacy which can undermine the computation. This flaw is connected (as is often the case) to regularization of divergent objects, and a consequent order-of-limits issue.
Still within the same computational framework, in section \ref{sec:twistone} I devise an alternative approach which aims at recovering the three-point function as a limit of its extension to the case where one of the twist-one operators possesses spin. This is an observable that I believe can be more safely computed with the method mentioned above, because its reduced symmetry provides some additional internal consistency checks, which the original three-point function is blind to. Indeed, such an analysis detects the presence of missing contributions, which I argue originate from the order-of-limits ambiguity pointed out above.
I isolate potentially dangerous contributions, which the method overlooks, determine them independently and re-incorporate them in my final estimate of the structure constant \eqref{eq:finalnorm}. This process relies on some assumptions, which I spell out in great detail, so to trace the origin of each contribution.

This result provides a successful test of the prediction \eqref{eq:prediction} derived assuming the conjecture on the matrix model description for the latitude Wilson loop. It also supplies some  two-loop novel results for structure contants of three-point functions of twist-one operators, two of which chiral primary and one un-protected (up to spin 6). These can be found in \eqref{eq:finalj} and table \ref{tab:3ptj}, along with the special case of three chiral primary operators \eqref{eq:finalnorm}, where I again stress that I find a result in disagreement with a previous computation \cite{Young:2014lka}.

\vskip 10pt

The connection that I present here, between a three-point function of local operators in ABJM and a Wilson loop expectation value computed with a matrix model, does not seem to be a common occurrence, but rather a very special and technical fact that is unlikely to be applicable to a variety of other contexts. 

Let me also stress that interesting connections between correlation functions of local operators and Wilson loop matrix models have emerged in various contexts in ${\cal N}=4$ SYM, when considering local operators inserted on Wilson lines, where a direct link with localization results can be established \cite{Giombi:2018qox,Giombi:2018hsx}.
In addition to such a connection, these Wilson line settings are also interesting as an instance of $AdS_2$ holography at strong coupling \cite{Giombi:2017cqn,Beccaria:2017rbe,Beccaria:2019dws} and to provide an interplay with integrability \cite{Kiryu:2018phb,Cavaglia:2018lxi}, and in particular the hexagon program for correlation functions \cite{Basso:2015zoa,Eden:2016xvg,Fleury:2016ykk}.
Speaking of which, the publication, \cite{Bargheer:2019kxb,Bargheer:2019exp} explored the computation of correlation functions of certain operators in ${\cal N}=4$ SYM that also reduce to matrix models.

In ABJM such endeavours relating correlation functions to matrix models have been mostly confined to topological sectors of the theory \cite{Dedushenko:2017avn}
or again to local operators on Wilson lines \cite{Bianchi:2017ozk,Bianchi:2018scb,Bianchi:2020hsz}, especially the relevant two-point correlators for the Bremsstrahlung function and its exact evaluation \cite{Correa:2012at}.
 
The computation presented here differs from those, and though it proposes a connection between a three-point function and a matrix model computing the expectation value of a Wilson loop, this does not directly emerge as a result of the correlation function being taken on the Wilson line and of localization being applied to them.
Rather, it is based on an analogy derived at the level of the Feynman diagrammatic expansions of two seemingly unrelated objects.

As mentioned above, integrability has recently been playing a major role in the understanding of correlation functions of ${\cal N}=4$ SYM, thanks to the hexagon approach.
Part of the motivation of this work is directed to provide some perturbative data that might turn helpful, should an extension of the hexagon program to ABJM be undertaken.
This is the case for the two-loop results that I supply here, especially on the ABJM slice $N_1=N_2$ where integrability is expected, rather than in the parity breaking ABJ extension where it could be spoiled.

The rest of the note explains how the various steps of the computation summarized in this Introduction were performed. Despite my efforts to keep the exposition uncluttered, the subject is unfortunately a bit technical, but I think it is worth spelling out some computation in greater detail. In particular, all the shaky points of my derivation are exposed, for future reference.

\section{Matrix model}\label{sec:latitude}

Supersymmetric Wilson loops can be defined in the ABJM model \cite{Aharony:2008ug,Aharony:2008gk}, which are in principle amenable of an exact computation via localization \cite{Pestun:2016zxk}.
This program has been accomplished extensively for the 1/6 \cite{Berenstein:2008dc,Drukker:2008zx,Chen:2008bp,Rey:2008bh} and 1/2 \cite{Drukker:2009hy} operators associated to a circular contour on the great circle of $S^3$ \cite{Kapustin:2009kz,Marino:2009jd,Drukker:2010nc}.
Supersymmetry can be conserved, albeit in a reduced amount, moving away from the great circle \cite{Cardinali:2012ru}. Among this class of operators a simple representative is the so-called latitude Wilson loop \cite{Bianchi:2014laa}, which is named after its contour being a circle at a certain azimuthal angle from the equator of $S^3$, which I dub $\Gamma_n$ ($n$ standing for potential multiple winding).
Supersymmetry on the sphere imposes a specific form of its connection, which depends on a deformation parameter $\nu$ (related to the azimuthal angle) and 
\begin{equation}\label{eq:latitude}
W_B = \frac{1}{ {\rm dim}(R)}\, \Tr_R\, \mathrm{P}\, \exp \left\{-i\oint_{\Gamma_n} d\tau \left(A_{\mu}\, \dot x^{\mu}-\frac{2 \pi i}{k}\, |\dot x|\, M^{A}_{\ \ B}\, Y^{B}\, \bar Y_{A}\right) \right\}
\end{equation}
where the matrix describing the coupling to the $(Y_I, \bar{Y}^I)$ scalars reads
\begin{equation}\label{eq:M} 
 \mbox{\small $\! M_{J}^{\ I}=\left(\!\!
\begin{array}{cccc}
 - \nu  & e^{-i \tau } \sqrt{1-\nu ^2} & 0 & 0 \\
e^{i \tau }  \sqrt{1-\nu ^2}  & \nu  & 0 & 0 \\
 0 & 0 & -1 & 0 \\
 0 & 0 & 0 & 1 \\
\end{array}
\right)$ }
\end{equation}
The Wilson loop described above possesses a $U(N_1)$ valued connection and for the rest of the paper I will consider the trace in the fundamental representation. An analogous version exists for the second gauge group $U(N_2)$, but here I will crucially focus on the $U(N_1)$ connection.
Moreover, I will restrict to the so-called bosonic loop, where the connection only contains bosonic blocks. A fermionic version constructed out of a supermatrix-valued connection exists as well, but I will not delve into that. This contains a coupling to the fermionic superpartners of the scalars $\psi_A$ and $\bar \psi^A$, where $A=1,\dots 4$. 
I refer to the appendix of \cite{Bianchi:2017svd}, for instance, for the explicit action of the model I am using.

Part of the interest in such a Wilson loop arises because of its relation to two-point functions of local operators on the Wilson loop, whose norm is related in turn to the energy emitted by an infinitesimally accelerating quark, the Bremsstrahlung function.
An exact knowledge of the latitude Wilson loop expectation value grants therefore a full knowledge of such a function, though in practice this step can be bypassed in such a way that the 1/6-BPS operator expectation value on the circle suffices \cite{Lewkowycz:2013laa,Bianchi:2014laa}.
Still, in \cite{Bianchi:2018bke} a matrix model computing the latitude Wilson loop expectation value was conjectured, with no direct proof using localization.
The proposal reads explicitly 
\begin{align} \label{eq:matrixlat}
\langle W_B^n(\nu) \rangle  &= \left\langle \frac{1}{N_1} \sum_{1\leq i\leq N_1} e^{2\pi\, n\, \sqrt{\nu}\, \lambda_{i}} \right\rangle
\end{align}
where $n$ stands for the winding number of the Wilson loop around its circular contour. The reason why I introduced multiple winding here is that it grants an additional probe for distinguishing perturbative contributions, that I will use in the next section.
The average is evaluated and normalized using the matrix model partition function
\begin{align} \label{eq:partitionf}
&  Z = \int \prod_{a=1}^{N_1}d\lambda _{a} \ e^{i\pi k\lambda_{a}^{2}}\prod_{b=1}^{N_2}d\mu_{b} \ e^{-i\pi k\mu_{b}^{2}}  \\
&~~ \times 
 \frac{\displaystyle\prod_{a<b}^{N_1}\sinh \sqrt{\nu} \pi (\lambda_{a}-\lambda _{b})\sinh\frac{ \pi (\lambda_{a}-\lambda _{b})}{\sqrt{\nu}} \prod_{a<b}^{N_2}\sinh\sqrt{\nu} \pi (\mu_{a}-\mu_{b}) \sinh\frac{\pi (\mu_{a}-\mu_{b})}{\sqrt{\nu}} }{\displaystyle\prod_{a=1}^{N_1}\prod_{b=1}^{N_2}\cosh\sqrt{\nu} \pi (\lambda _{a}-\mu_{b}) \cosh\frac{\pi (\lambda _{a}-\mu_{b})}{\sqrt{\nu}} } \nonumber
\end{align}
Despite a number of perturbative checks \cite{Bianchi:2018scb,Aguilera-Damia:2018bam,David:2019lhr,Medina-Rincon:2019bcc}, \eqref{eq:matrixlat} remains a conjecture and henceforth amenable of further tests, which is part of the purpose of this paper.

\section{A three-point function from the matrix model}\label{sec:MM23pt}

\begin{figure}
\centering
\includegraphics[width=3.5cm]{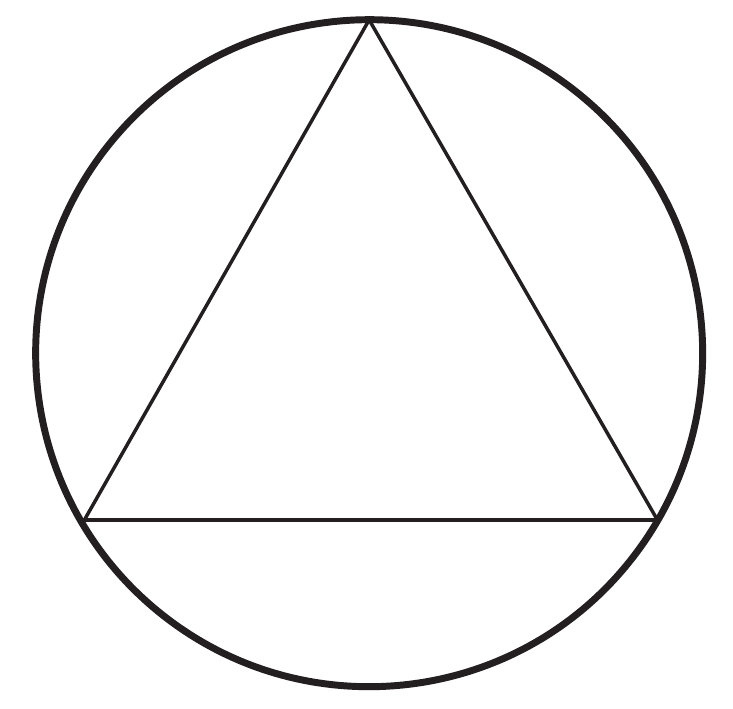}
\caption{Scalar triangle diagram contributing to the latitude Wilson loop.}\label{fig:triangle}
\end{figure}

The peculiar aspect of such a Wilson loop on which I want to focus in this note consists in the presence, in its weak coupling perturbative expansion, of a scalar triangle diagram, depicted in Figure \ref{fig:triangle}. 
This contribution first appears at three loops, it is purely imaginary, as expected at odd loop order, and, contrary to other such terms, it does not originate from a non-trivial framing \cite{Calugareanu1961,Witten:1988hf} of the Wilson loop.

This kind of diagrams is a special feature of supersymmetric Wilson loops in ABJM, which contain a coupling to a bi-scalar operator. Still, in the ordinary case of the 1/6-BPS Wilson loop on the great circle this kind of contribution vanishes identically because the corresponding trace of the product of three matrices governing this coupling (same as \eqref{eq:M} with $\nu=1$) vanishes identically.
A similar diagram is non-vanishing in the case of the 1/2-BPS Wilson loop on the great circle. However, the perturbative expansion of that object is complicated by the supermatrix structure of its connection and in particular the coupling to fermions \cite{Bianchi:2013pva,Bianchi:2013rma,Griguolo:2013sma}. Since most of the arguments that I will use momentarily apply to the perturbative expansion of Wilson loop, I will only deal with the bosonic operators, for which it is much better developed.

Perturbatively, the highlighted scalar triangle contribution resembles the (integrated over the Wilson loop parameters) three-point function of scalar bi-linear operators of the form $M_A^{\phantom{A}B}(\tau) Y^A \bar{Y}_B$.
More precisely, I will focus on a particular flavor choice, so as to minimize contractions, such as
\begin{equation}\label{eq:operators}
O_1(x_1) = \Tr (Y^1 \bar Y_2) \quad,\quad O_2(x_2) = \Tr (Y^2 \bar Y_3) \quad,\quad O_3(x_3) = \Tr (Y^3 \bar Y_1)
\end{equation}
and consider their three-point function
\begin{equation}\label{eq:3ptstruc}
\left\langle O_1(x_1) \, O_2(x_2) \,  O_3(x_3) \right\rangle = \frac{\tilde{\cal C}}{|x_{12}|\, |x_{23}|\, |x_{13}|} \qquad,\qquad x_{ij}\equiv x_i - x_j
\end{equation}
where the space-time structure is fixed by conformal invariance and the only free parameter is the (un-normalized) structure constant $\tilde{\cal C}$, which is a function of the coupling $k^{-1}$ and the gauge group ranks $N_1$ and $N_2$.
I will adopt the notation without tilde for the structure constant normalized by the two-point function of the operators.

Therefore, in a sense that I will make sharper momentarily, the latitude Wilson loop expectation value contains information of three-point functions of scalar operators of the form $Y\bar{Y}$ in the ABJM model. The crucial question is whether it is possible to extract sensible data from this. Potentially, this could be extremely rewarding as it could grant all-order results for some three-point functions.

\subsection{Colorful considerations at two loops}

Two kinds of obstructions arise against carrying out such a program. First, in the Wilson loop as computed exactly by the matrix model, one only obtains the total expectation value, and not individual contributions, which can be directly associated to certain diagrams. Therefore it might be completely impossible to disentangle the relevant parts to the three-point function from those pertaining to other perturbative contributions to the Wilson loop expectation value.

\begin{figure}
\centering
\includegraphics[width=7.5cm]{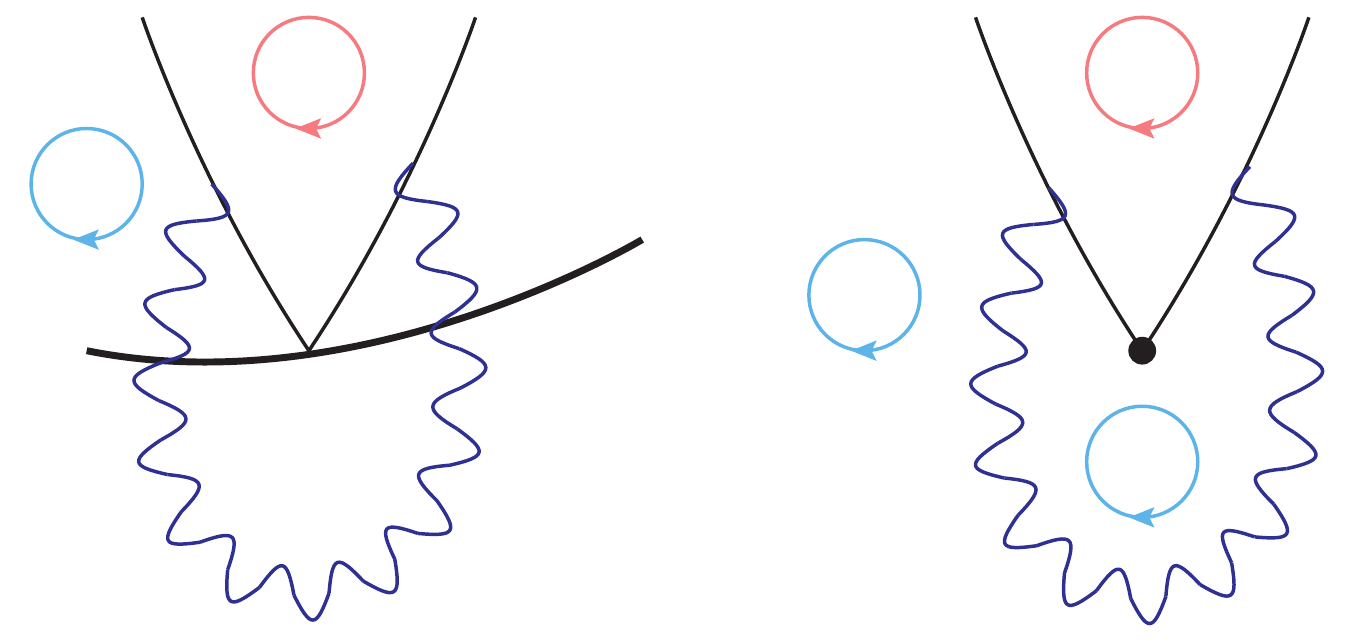}
\caption{Cartoon of the different behaviour in color space between the insertion of an operator on a Wilson line and the color trace of an operator such as in \eqref{eq:operators}. In the example the former is not color leading, whereas the latter is. Color loops of the two different gauge groups are indicated in blue and red. For instance, the diagram on the left gives rise to a $N_1$ subleading contribution (and indeed it is graphically non-planar), whereas the graph on the right is leading in $N_1$.}\label{fig:color}
\end{figure}

On the other hand, the perturbative expansion of triangle diagrams in the matrix model and the computation of the corresponding three-point function are only qualitatively similar, but actually rather different. The color structure, in particular, is completely distinct: while in a three-point function computation one faces the contraction of three single trace operators of the schematic form $\Tr(Y\bar{Y})$, in the bosonic Wilson loop computation a trace over three scalar bi-linear insertions is taken, such as $\Tr(Y\bar{Y}(\tau_1)Y\bar{Y}(\tau_2)Y\bar{Y}(\tau_3))$.
In general, their perturbative expansions are different as for instance some diagrams in the first case are color leading, whereas they are sub-leading in the latter, as illustrated graphically in a simple example in Figure \ref{fig:color}.

Let me limit the description now to the first non-trivial quantum corrections to three-point functions, which appears at two loops. The corresponding graphs would contribute to the five-loop expectation value of the supersymmetric Wilson loop.

Using a similar graphical representation (or an explicit computation) it can be realized that there is a subclass of diagrams for which the color structures in the Wilson loop and the three-point function computations are simply proportional (with the same constant of proportionality, dictated by the lowest order computation). 
This is the situation, when the leading $N_2$ color contribution is retained in both calculations. This corresponds graphically to planar contributions which lie {\it internally}, so to say, to the scalar triangle.
This graphical argument, basically guarantees that in this limit the $N_2$ color algebra is the same for the two computations and only the $N_1$ contractions are performed differently, from which the proportionality emerges.

Further, the limit also restricts considerably the number of allowed graphs, since the bi-fundamental nature of the ABJM matter fields 
In particular, the limit forces quantum corrections to consists only of $U(N_2)$ gauge vectors, certain terms in the Yukawa vertices and, importantly, they cap the number of possible sextic scalar interaction vertices to only one. 
An exception to the rule of thumb for the proportionality of color structures, is in fact produced by a single class of diagrams involving the sextic interaction, that does not obey the same color scaling as all other graphs in this limit. But let me postpone this discussion to section \ref{sec:conjecture}, in order not to interrupt the line of reasoning at this point.
Therefore I will consider such a color limit from now on and ascertain if any mileage can be gotten from this setting.

The second substantial difference between the Wilson loop and three-point function computations lies in the fact that the relevant operators are substantially different. 
For the ordinary case of chiral primary operators in ABJM, these are given by a usually position independent (although one may try and construct more general operators, resembling the superprotected of \cite{Drukker:2009sf}) symmetric and traceless flavor matrix $\tilde M$ (I reserve the notation $M$ for the coupling matrix of scalars in the Wilson loop connection \eqref{eq:M}). Up to a normalization factor, these can be chosen to be operators of the form $Y^A\bar{Y}_B$, where $A\neq B$.

The bi-linears appearing in the Wilson loop are instead coupled to a position dependent matrix (to ensure global supersymmetry of the Wilson loop connection), which is still traceless.
Tracelessness, together with the flavor symmetry of ABJM interactions, implies that triangle graphs in the Wilson loop expectation value can only depend on a single flavor matrix contraction $\Tr(M(\tau_1)M(\tau_2)M(\tau_3)) = -\Tr(M(\tau_1)M(\tau_3)M(\tau_2))$ (where the equality stems from the particular structure of the coupling matrix $M$). 
As a consequence, these diagrams possess the same perturbative expansion and are therefore proportional to (still assuming the large $N_2$ color limit) those of an ordinary three-point function calculation, where the scalar flavors of the chiral primary operators are chosen so as to allow for a unique contraction, which is precisely the case for the three-point functions \eqref{eq:3ptstruc} under exam.

The conclusion of this analysis is that in the large $N_2$ limit the perturbative expansion of the two-loop three-point function \eqref{eq:3ptstruc} is proportional to that of the corresponding scalar triangle graphs in the Wilson loop perturbative expansion at five loops. 
The constant of proportionality amounts to a color factor and a number produced by the integration over the Wilson loop parameters. 
The latter is computed straightforwardly using the fact that the structure of the triangle diagrams is fixed by conformal symmetry, as it coincides with a three-point function of primary operators \eqref{eq:3ptstruc}, and that the trace of three $M$ coupling matrices \eqref{eq:M} is precisely of the form to cancel such a position dependence (when the operators are located on a circular contour). The correlator then becomes topological (or position independent), and the contour integration produces a straightforward factor
\begin{equation}
\int_0^{2\pi} \int_0^{\tau_1} \int_0^{\tau_2} d\tau_1\, d\tau_2\, d\tau_3 = \frac{4}{3}\, \pi^3
\end{equation}
Still, it remains to be understood whether the contributions to such triangle graphs can be isolated in the matrix model average, in order to derive sensible information on them.

\subsection{Color limit on the Wilson loop}

First, the color limit $N_2\gg N_1\gg 1$ has to be implemented.
At the weak coupling perturbative level this can be realized by simply restricting to certain classes of diagrams with a maximal amount of insertion points on the contour.

Let me clarify this.
The leading $N_2$ color contribution for a Wilson loop with a $U(N_1)$ gauge connection reads $N_1 N_2^l$ at $l$ loops, according to the normalization \eqref{eq:latitude} of the Wilson loop operator.
The fact that only a single power of $N_1$ is retained, means that only diagrams with two insertion points are allowed, as any additional point would necessarily create a $U(N_1)$ color loop.
This restricts the analysis to a single gluon or scalar bi-linear exchange, where all their loop corrections with only $N_2$ color powers are allowed. 
In \cite{Bianchi:2016rub} Matias Leoni and I explained how this information suffices to solve the problem completely, and how to extract these all-order corrections from the matrix model result for the 1/6 bosonic Wislon loop on the great circle.

The result of such an analysis reads (the blobs stand for all-order corrections in $\frac{N_2}{k}$)
\begin{align}\label{eq:leading}
\raisebox{-.4cm}{\includegraphics[width=2.cm]{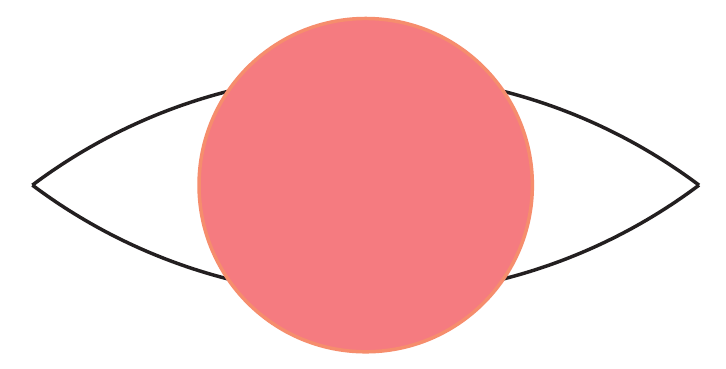}} &= \frac{\sin \frac{\pi N_2}{k}}{\frac{\pi N_2}{k}}\raisebox{-.4cm}{\includegraphics[width=2.cm]{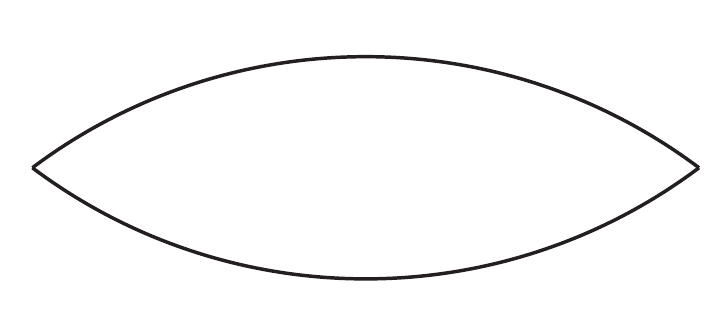}}\\
\raisebox{-.4cm}{\includegraphics[width=2.cm]{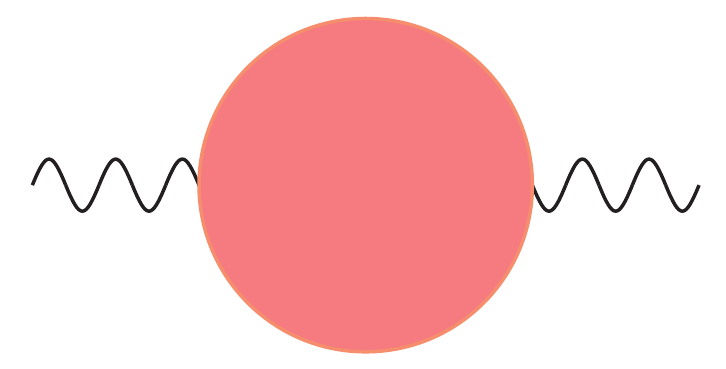}} &= \cos \frac{\pi N_2}{k} \raisebox{-0.025cm}{\includegraphics[width=2.cm]{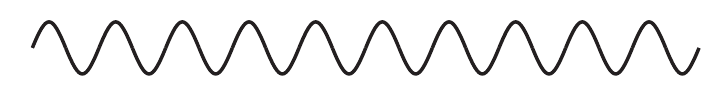}} + \frac{\sin \frac{\pi N_2}{k}}{\frac{\pi N_2}{k}} \raisebox{-.4cm}{\includegraphics[width=2.cm]{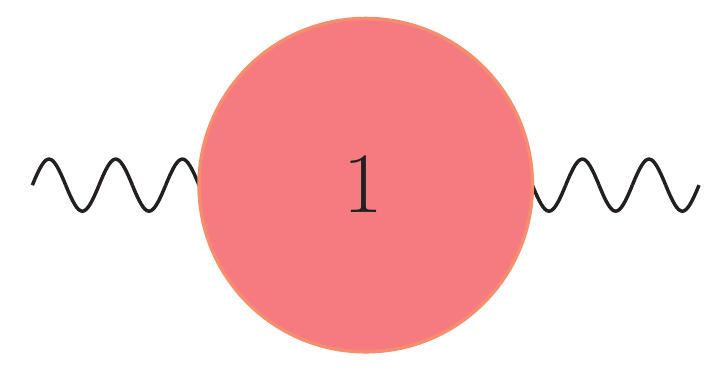}}\label{eq:leading2}
\end{align}
in terms of the tree-level bi-scalar and gluon exchanges and the one-loop gluon propagator, which are the lowest order independent structures.
Incidentally, the former contribution coincides with the all-order two-point function of scalar operators of the form $\Tr(Y^A \bar{Y}_B) - \text{trace}$, which are relevant for the three-point function computation as well, since eventually one is after the structure constant normalized by the two-point functions of the relevant operators. The above results provide such a tool, to all orders, in the leading $N_2$ limit. Also, their knowledge suffices to infer the whole $N_2$ leading limit of an extremal correlation function, as I explain in section \ref{sec:extremal}.

\begin{figure}
\centering
 \begin{subfigure}{2.6cm}
  \centering
  \includegraphics[width=2.5cm]{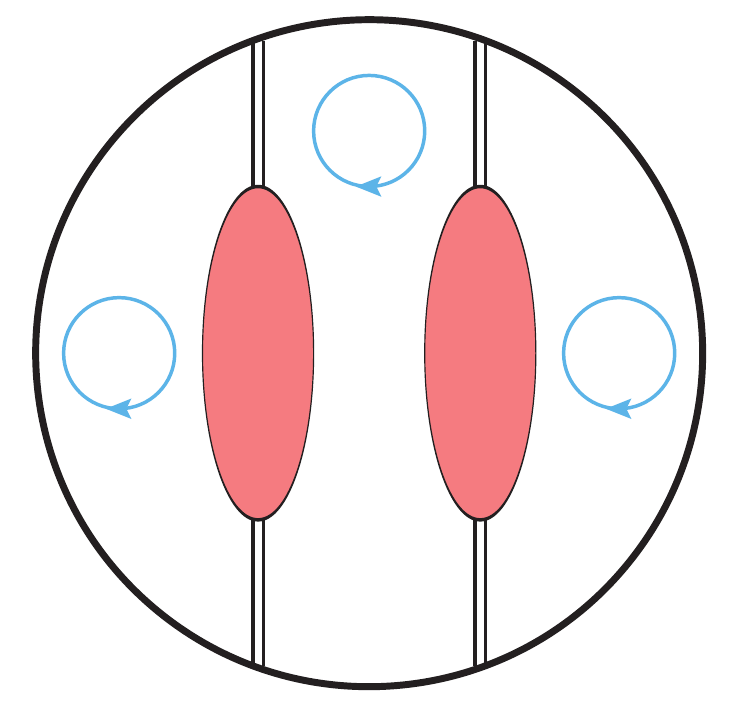}
  \caption{}
  \label{fig:3Ldiagramsa}
  \end{subfigure}  \quad
 \begin{subfigure}{2.6cm}
  \centering
  \includegraphics[width=2.5cm]{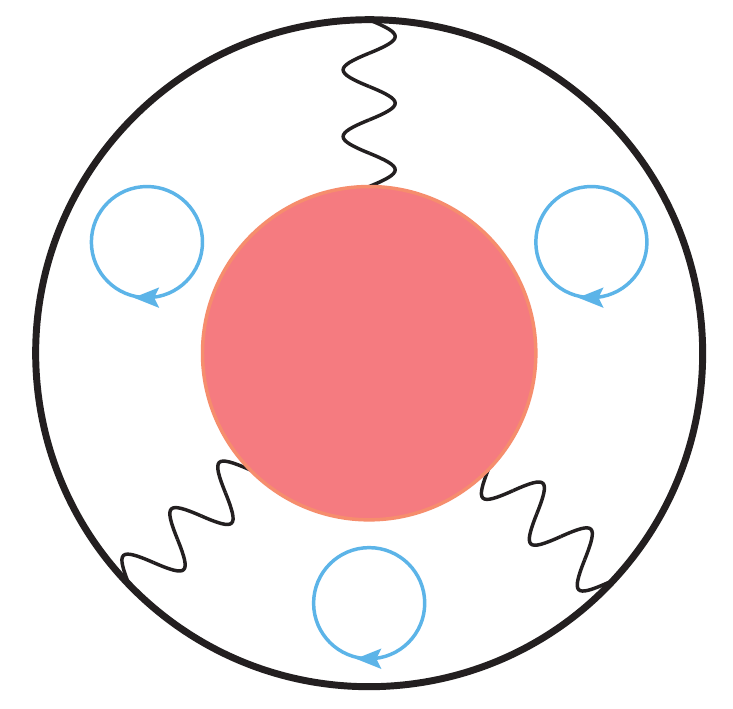}
  \caption{}
  \label{fig:3Ldiagramsb}
  \end{subfigure}  \quad
   \begin{subfigure}{2.6cm}
  \centering
  \includegraphics[width=2.5cm]{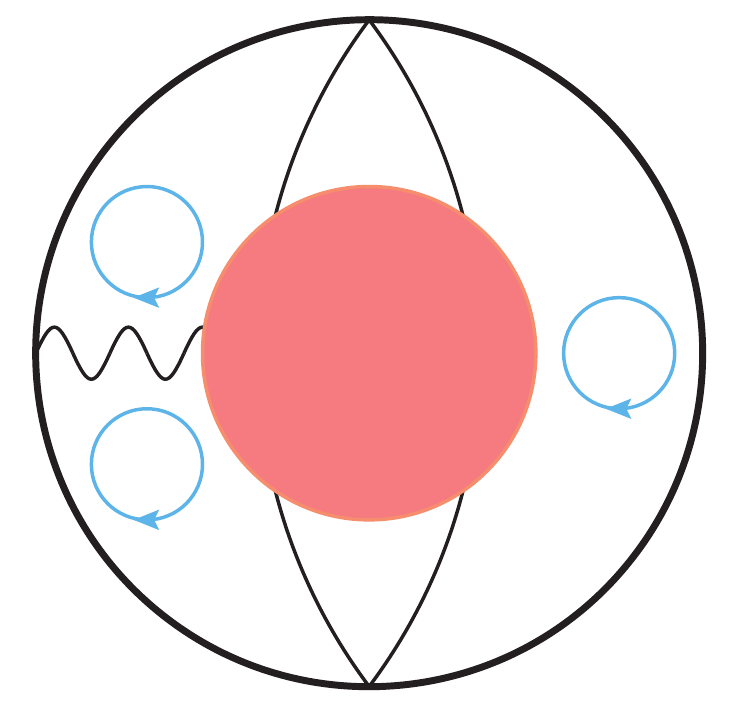}
  \caption{}
  \label{fig:3Ldiagramsc}
  \end{subfigure}  \quad
 \begin{subfigure}{2.6cm}
  \centering
  \includegraphics[width=2.5cm]{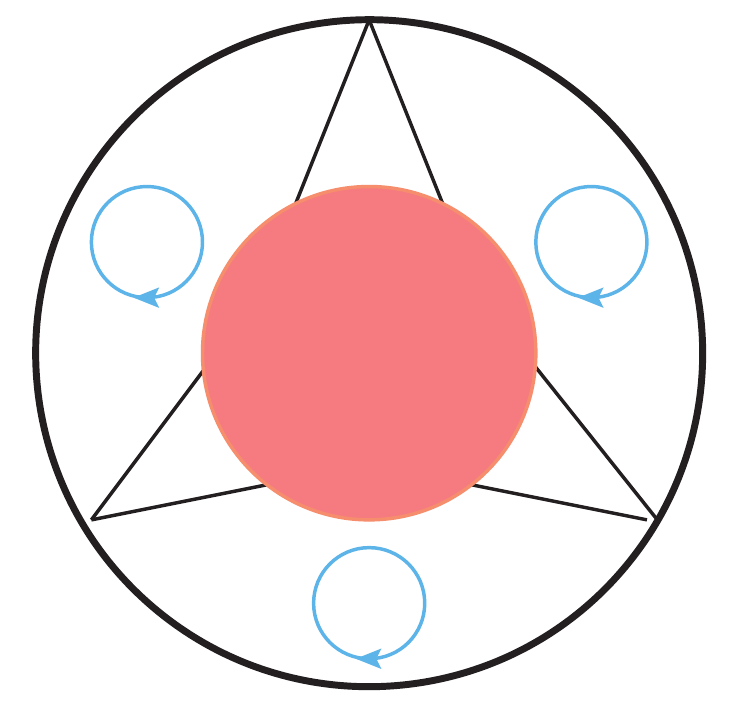}
  \caption{}
  \label{fig:3Ldiagramsd}
  \end{subfigure}  \quad
\begin{subfigure}{2.6cm}
  \centering
  \includegraphics[width=2.5cm]{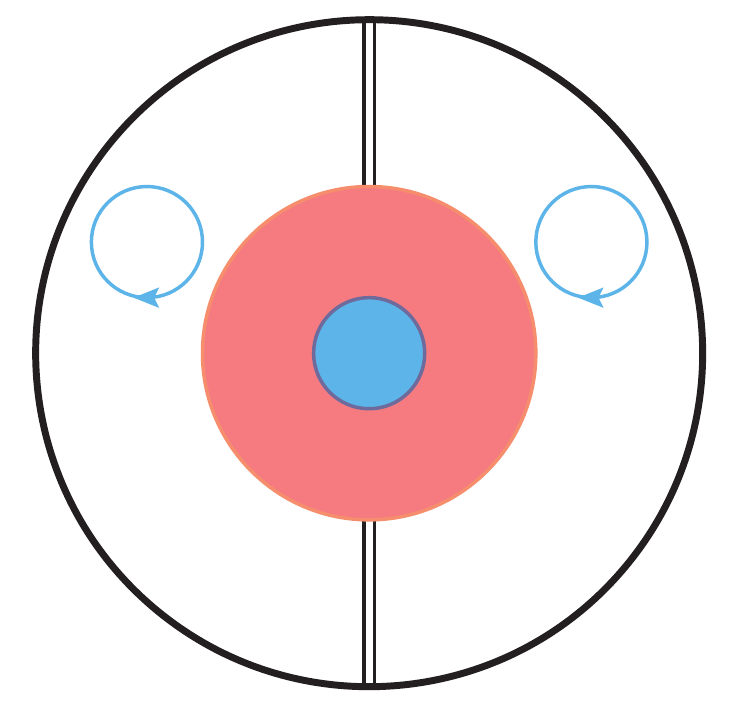}
  \caption{}\label{fig:3Ldiagramse}
  \end{subfigure}  \quad
\caption{Diagrams contributing to the latitude Wilson loop expectation value at sub-leading $N_2$ order, which is $N_1^2 N_2^{l-2}$ at $l$ loops. The blue circles represent loops in $N_1$ color space, giving rise to $N_1$ powers. To this order in the color limit, only three $N_1$ loops are allowed, which after normalization give rise to the desired $N_1^2$ factor. Solid lines stand for scalars, whereas wavy for gluons. Red bubbles represent all quantum corrections with maximal $N_2$ power at each perturbative order. The double line in diagrams $(a)$ and $(b)$ represent the combined exchange of a gluon and a scalar pair, from the connection \eqref{eq:latitude}.}\label{fig:3Ldiagrams}
\end{figure}

The triangle graphs I am mostly interested in in this note, do not lie in the class with maximal $N_2$ power.
Rather, they possess at most a $N_2^{l-1}$ power. This belongs to the next-to-leading correction at $N_2\to\infty$.
I find that the weak coupling expansion of the matrix model up to this order in colors, in powers of $\frac{N_2}{k}$ can be re-summed according to
\begin{align}
\langle W_B^n(\nu) \rangle &= \frac{N_1}{k} \pi n^2 \left(  \frac{\nu ^2+1}{2} \sin\frac{\pi  N_2}{k}
+ i  \nu \, \cos\frac{\pi  N_2}{k} \right)
+ \nonumber\\ &
\frac{\pi ^2 n^2 N_1^2}{24 k^2} 
\left[
4 i \nu  \left(\left(\left(\nu ^2+1\right) n^2-2\right) \sin\frac{2 \pi  N_2}{k} + 6 \sin \frac{\pi  N_2}{k}\right) \right.\nonumber\\& 
+ 4 \left(\nu ^2+1\right) \cos\frac{\pi  N_2}{k} \left(2 \cos\frac{\pi  N_2}{k}-3 \right)
+\left(\nu ^2-1\right)^2 n^2
\nonumber\\& \left.
- \left(\nu ^4+6 \nu ^2+1\right) n^2 \cos\frac{2 \pi  N_2}{k} \right]
 +O\left(N_1^3\right) + \text{color subleading}
\end{align}
The leading term in $N_2$ can be derived precisely using the results \eqref{eq:leading} and \eqref{eq:leading2}, whereas the next to leading piece is that I want to focus on mostly.
Pictorially, the relevant diagrams in the Wilson loop expansion to this particular order in colors can have at most four insertion points.
In practice, these correspond to the generic classes of Figure \ref{fig:3Ldiagrams}, where I am understanding all possible corrections in the blobs, in particular not only those given by 1PI contributions.
Diagrams with three or four insertion points already possess at least a power $N_1^2$, therefore only additional $N_2$ color factors are allowed in their perturbative expansions at fixed $N_1^2 N_2^{l-2}$ order at $l$ loops.
On the other hand, diagrams with two insertion points, namely those considered in the leading $N_2$ factor above, have to be considered with their $N_1 N_2^{l-1}$ corrections (here I mean the correction to the gauge and bi-scalar two-point function at loop $l$, which produce diagrams of order $l+1$ and $l+2$, respectively), that is at next-to leading order in large $N_2$.

I recall that the main idea behind this reasoning  consists in eventually trying and singling out the particular contribution of scalar triangle graphs, which are relevant for the three-point function.
In order to achieve this I now use all the information I possess on the perturbative expansion of the latitude Wilson loop, concerning its dependence on the multiple winding parameter $n$ \cite{Bianchi:2016gpg}, framing $f$ and latitude parameter $\nu$ and their better understood limit onto the great circle 1/6-BPS operator. The framing number is eventually set to $f=\nu$, as this is conjecturally the value relevant for the latitude matrix model result, however I keep it generic in the perturbative contributions to better separate them.

I start with the four insertion points diagrams. These are obtained by the superposition of double exchanges of gluons and scalar bi-linears. In the leading $N_2$ limit, their individual contribution is known from \cite{Bianchi:2016rub}, by comparison to the un-deformed Wilson loop expectation value. The combination of two such exchanges can be handled in different ways. One consists in adding and subtracting a crossed exchange configuration which would be color sub-leading in itself, in order to produce a combination which factorizes into the product (or the square in this case) of two exchanges, which I know from \cite{Bianchi:2016rub}.

The artificially introduced crossed contribution vanishes if at least one of the exchanges has the Chern-Simons propagator structure, since no framing contribution is produced in this case and consequently the corresponding graph vanishes on a plane. In the case where two even exchanges are considered, the corresponding crossed configuration does not vanish. Nevertheless such a term is simple to compute, since the combined gluon and scalar bi-linear contribution (indicated with a double line) is topological (also in the latitude case) and consequently it can be evaluated straightforwardly.
The final result reads
\begin{align}\label{eq:doubleex}
\raisebox{-0.8cm}{\includegraphics[width=2.cm]{mixedColor}} + \text{perms} &= 
\frac{N_1^2 \pi^2}{12k^2} \left(\left(\nu ^2+1\right)^2 n^4 \sin^2\frac{\pi  N_2}{k} \right.\nonumber\\&\left.
- n^2 \left(2 n^2+1\right) \left(2f^2 \cos ^2\frac{\pi  N_2}{k}
-i f \left(\nu ^2+1\right) \sin\frac{2 \pi  N_2}{k} \right) \right)
\end{align}
Next I separate the analysis between even and odd perturbative orders.
I recall that the scalar triangle graphs start appearing at three-loops, therefore the scalar triangle (prospectively three-point function) even loop corrections appear at odd loops in the Wilson loop and vice versa.
From considering the 1/6-BPS Wilson loop where scalar triangle graphs are absent, 
after subtracting the double exchange contribution \eqref{eq:doubleex} (at $\nu=1$), the remaining diagrams contribute to the $N_2$ sub-leading part at odd orders
\begin{equation}\label{eq:diagodd}
\raisebox{-0.8cm}{\includegraphics[width=2.cm]{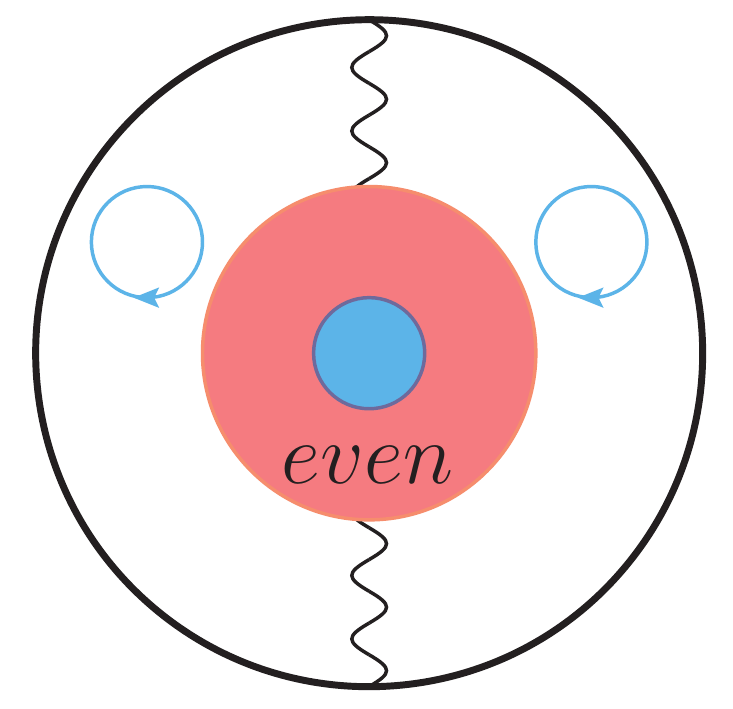}} + \raisebox{-0.8cm}{\includegraphics[width=2.cm]{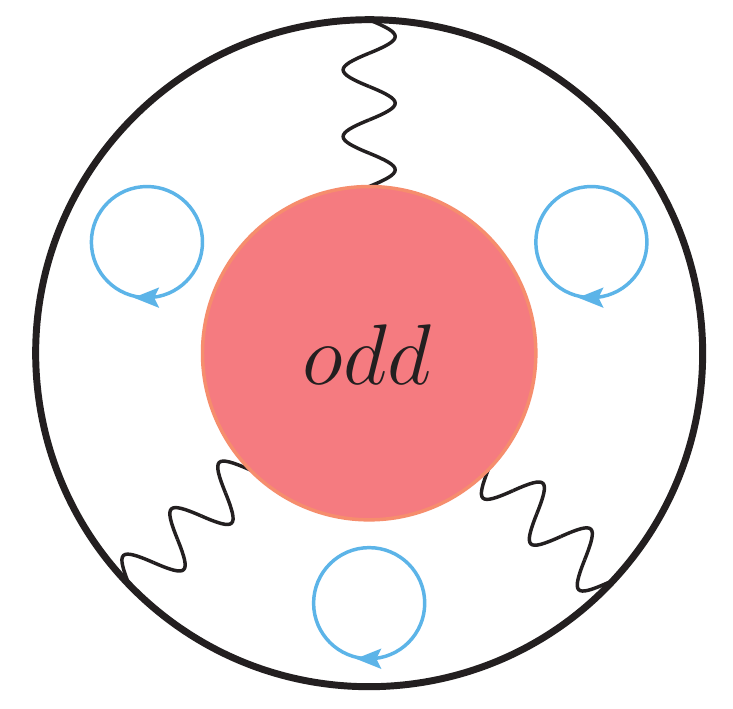}} + \raisebox{-0.8cm}{\includegraphics[width=2.cm]{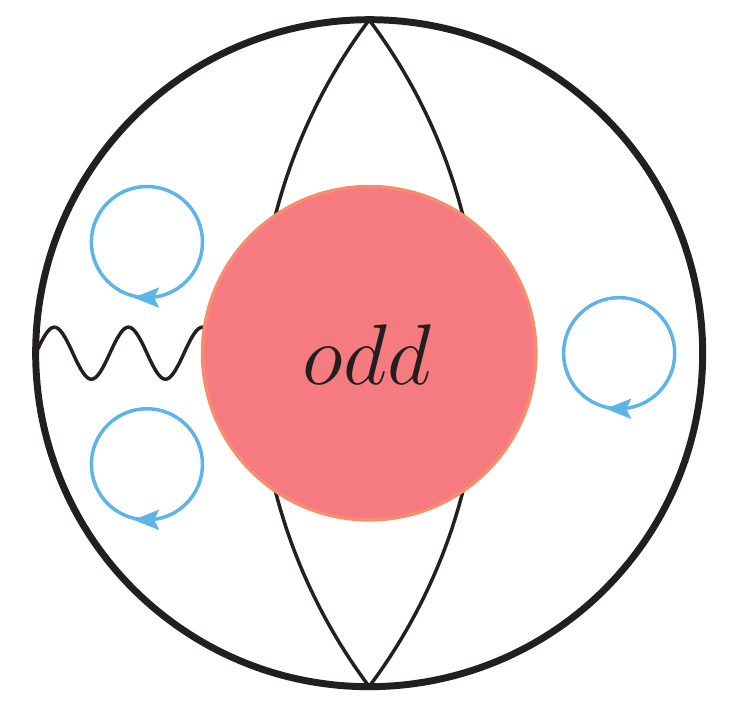}}\,\, \Bigg|_{\nu=1} =
\frac{ 2 i n^2 \pi^2 N_1^2 \sin^2\frac{\pi N_2}{2k} \sin\frac{\pi N_2}{k}}{k^2}
\end{equation}
My use of {\it even} and {\it odd} in the pictures could be a bit misleading. With those I mean the overall loop order of the whole corrected structure. 
Such diagrams are the same as those appearing in Figure \ref{fig:3Ldiagramsb}, \ref{fig:3Ldiagramsc} and \ref{fig:3Ldiagramse}, evaluated at $\nu=1$. Diagram \ref{fig:3Ldiagramse} reduces to the gluon exchange at odd orders, since the corresponding corrections to the bi-scalar exchange vanish.
At three loops the corresponding contribution \eqref{eq:diagodd} vanishes, meaning that a non-trivial cancellation occurs between the gluon exchange and the three-insertions diagrams \cite{Bianchi:2016yzj}.

In \cite{Bianchi:2018bke} it was argued that such a cancellation at three loops also applies to the latitude case.
It is therefore conceivable that the total net effect of these diagrams for the latitude keeps enjoying similar cancellations at higher orders as well, which in the un-deformed case led to \eqref{eq:diagodd}. 
In particular, turning on the latitude deformation presents two sources of $\nu$ corrections. The first is associated to the matrix model working conjecturally at framing $\nu$. The second arises from the coupling matrix $M$. The latter appears only in diagram \ref{fig:3Ldiagramsc}, however the $\nu$ dependence is associated with a $\tau$-dependent factor that at three loops makes the corresponding diagram insensitive to framing and therefore vanishing. Conjecturing that the same mechanism also occurs at higher loop orders, this source of $\nu$ corrections can be discarded.
This leaves with $\nu$ contributions stemming from framing.
This is the case for the gluon exchange which is proportional to the Gauss linking number, yielding a linear dependence on $\nu$, at framing $\nu$. A thorough analysis of the framing dependence of the remaining diagrams \ref{fig:3Ldiagramsb} and \ref{fig:3Ldiagramsc} is intricate and missing.
The only certain piece of information is that their framing contribution should not vanish for $\nu\to 1$, otherwise it could not account for \eqref{eq:diagodd}. 

I will now assume that the hole sum of diagrams \eqref{eq:diagodd} is linear in the framing number 
\begin{equation}\label{eq:odd}
\raisebox{-0.8cm}{\includegraphics[width=2.cm]{gaugeOddColorSub}} + \raisebox{-0.8cm}{\includegraphics[width=2.cm]{gauge-vertex-colorOdd}} + \raisebox{-0.8cm}{\includegraphics[width=2.cm]{scalar-gaugeColorOdd}}  \underset{\text{assumption}}{=} \frac{2 i \pi ^2 f n^2 N_1^2 \sin ^2\frac{\pi  N_2}{2 k} \sin \frac{\pi  N_2}{k}}{k^2}
\end{equation}
and explore the consequences.
Under this assumptions I subtract the odd loop contributions of \eqref{eq:odd} to the latitude matrix model expansion in the sub-leading large $N_2$ component and find the simple result
\begin{align}\label{eq:trifromMM}
\raisebox{-0.8cm}{\includegraphics[width=2.cm]{triangleColor}} &= -\frac{i\, \pi ^2\, \nu  \left(\nu ^2-1\right) n^2\, N_1^2\, \sin \frac{2 \pi  N_2}{k}}{12\, k^2}\nonumber\\&
= \frac{\sin \frac{2 \pi  N_2}{k}}{\frac{2\pi N_2}{k}}\, \times \raisebox{-0.8cm}{\includegraphics[width=2.cm]{triangle}}
\end{align}
where in the second line I have rewritten the expression factorizing the lowest order  triangle diagram, appearing at three loops in the Wilson loop expectation value.
The two-loop correction to the triangle structure in the color limit, which is the relevant piece of information for this work, reads
\begin{equation}\label{eq:trifromMM2}
\raisebox{-0.8cm}{\includegraphics[width=2.cm]{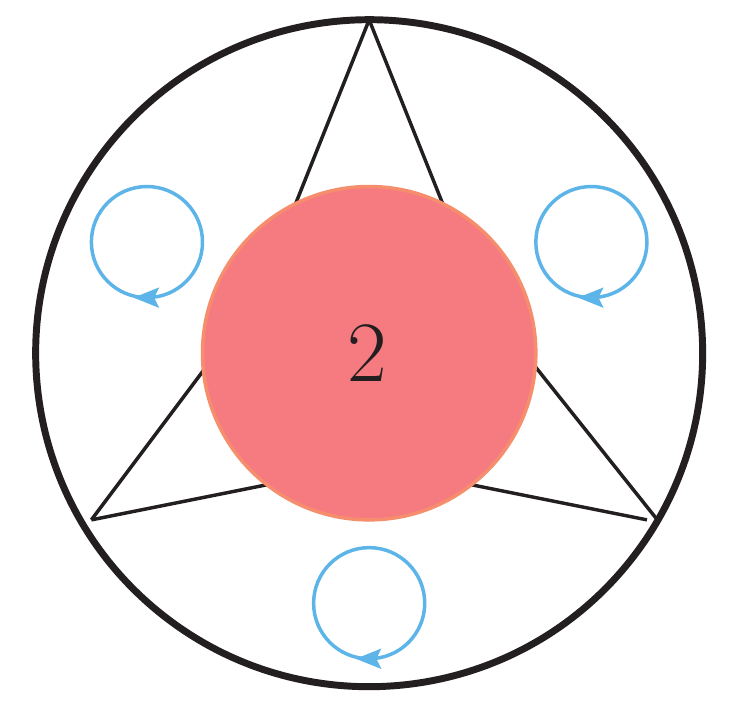}} 
= -\frac{2\pi^2}{3}\, N_2^2\, \times \raisebox{-0.8cm}{\includegraphics[width=2.cm]{triangle}}
\end{equation}
My observation here is that the $\nu$ dependence of this estimate coincides precisely with that expected from scalar triangle diagrams, where it is generated by the trace of three $M$ matrices.
This seems to hint at a self-consistency of the assumption I made on the diagrams \eqref{eq:odd}. Had they had a different dependence on $\nu$ than the one assumed, they would likely change the $\nu$ dependence of \eqref{eq:trifromMM}. They would do it certainly at three loops whereas I cannot say this with certainty at higher orders, where in principle they could also be proportional to $\nu(\nu^2-1)$. However it would seem unlikely to me that this precise behavior occurred at all orders but three loops.

In conclusion, under an assumption \eqref{eq:odd} on the $\nu$ dependence of the diagrams \ref{fig:3Ldiagramsb} and \ref{fig:3Ldiagramsc}, I infer that the even order corrections of the triangle diagrams in the matrix model (which appear as odd loop order corrections in the Wilson loop expectation value), with maximal $N_2$ power loop-by-loop, read \eqref{eq:trifromMM}.

Before relating this to the three-point function of chiral primary operators of length 2, let me also examine the even perturbative orders of the latitude matrix model expansion at sub-leading order in large $N_2$.

In this case, apart from the double exchange \eqref{eq:doubleex}, the contributions of diagrams \ref{fig:3Ldiagramsb}, \ref{fig:3Ldiagramsc} and \ref{fig:3Ldiagramse} could have independent values, or at least I do not have any means of restricting their form at the moment. From lower order perturbative computations carried out for various purposes (where similar structures appear as intermediate steps), I checked that divergences may also be generated by the individual contributions separately, which require a mechanism to cancel, which can involve mixing of singularities coming from internal integrations and those over the Wilson loop parameters.
Still, at even loops their $\nu$ dependence in the latitude case is completely fixed by the $\Tr(M(\tau_1)M(\tau_2))$ structure and can occur only in diagrams \ref{fig:3Ldiagramsc} and \ref{fig:3Ldiagramse}. For them the combined expansion reads (always restricting to the color limit)
\begin{equation}
\raisebox{-0.8cm}{\includegraphics[width=2.cm]{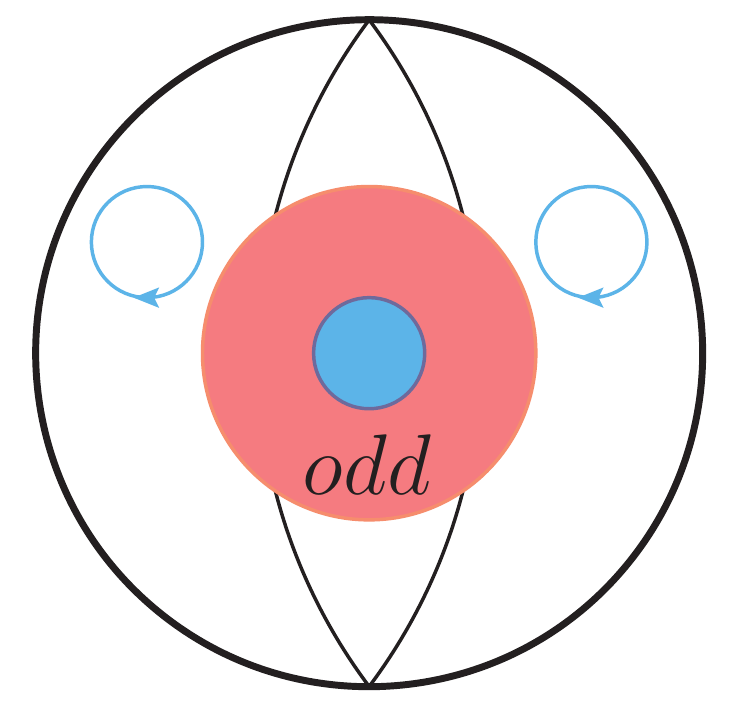}} + \raisebox{-0.8cm}{\includegraphics[width=2.cm]{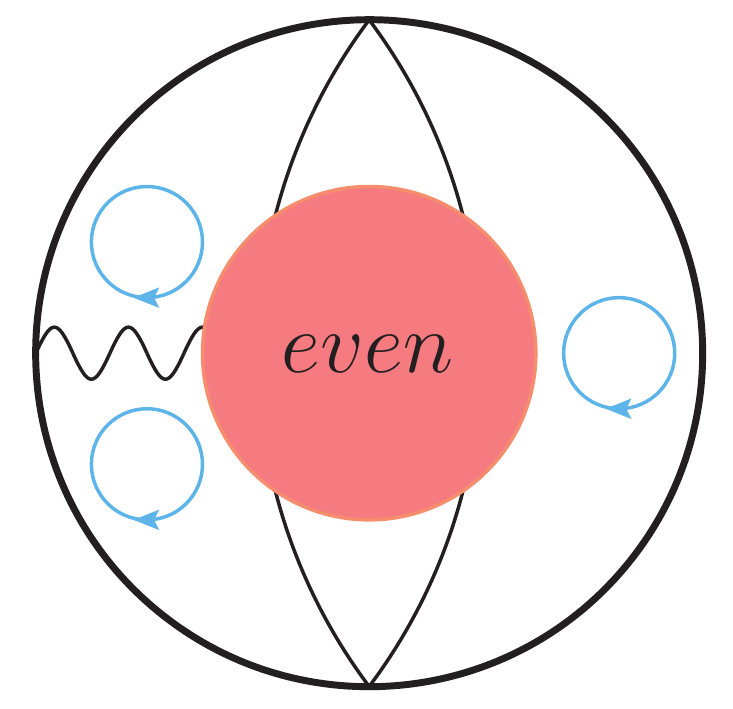}}
=-\frac{\pi ^2 \left(\nu ^2-1\right) n^2 N_1^2 \sin ^2\frac{\pi  N_2}{2 k} \cos \frac{\pi  N_2}{k}}{k^2} + \text{$\nu$-independent }
\end{equation}
The dependence on $\nu$ via the factor $\nu ^2-1$ is that contained explicitly in the coupling matrices $\Tr(M(\tau_1)M(\tau_2))$ product. At two loops only the first diagram above contributes, whereas I expect in general their sum to give rise to the $\nu$ dependence at higher orders, in the color limit.

\subsection{Extracting the two-loop color limit of the three-point function}\label{sec:conjecture}

Equipped with the results of the previous section and in particular \eqref{eq:trifromMM2}, I can now extract a prediction for the color component of the three-point function \eqref{eq:3ptstruc} at two loops.
Before that, let me recall that in the color limit, the scalar triangle diagrams appearing in the Wilson loop perturbative expansion and the corresponding graphs of the three-point function at two loops are related by a common proportionality factor.
However there is a class of anomalous diagrams, and only one at two-loop order, for which the proportionality factor is different from all others and therefore has to be treated distinctly.

\begin{figure}
\centering
\includegraphics[width=3.5cm]{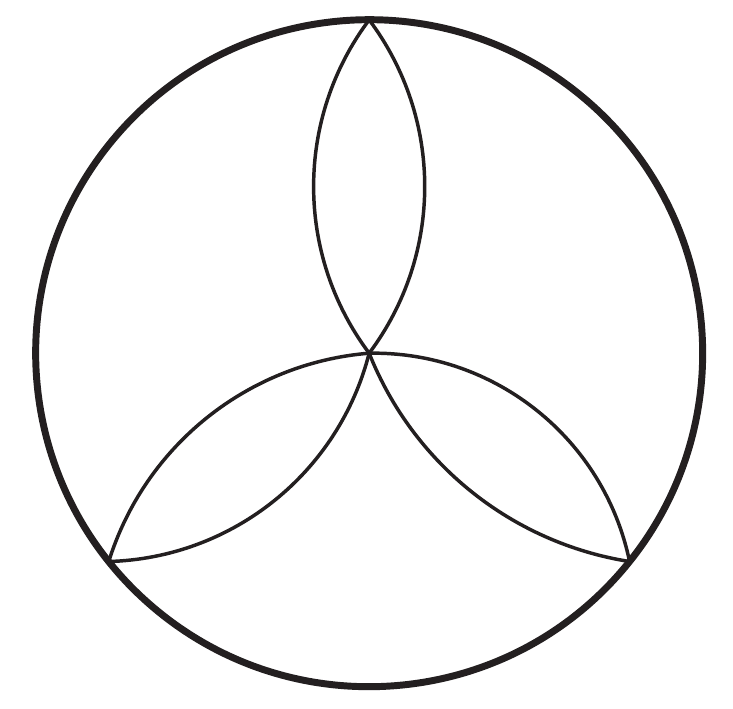}
\caption{Clover diagram contributing to the expectation value of the latitude Wilson loop at 5 loops.}\label{fig:clover}
\end{figure}

This is what I would refer to as a {\it clover} diagram, which is depicted at the lowest perturbative order in Figure \ref{fig:clover}.
An analogous diagram also plays a somewhat distinct role in the computation of three-point functions in section \ref{sec:clover}.
The contribution from the clover graph can be derived from the explicit computation in section \ref{sec:clover} and formula \eqref{eq:cloverint} (adapting the overall factors suitably). Anticipating that result the clover contribution in the color limit reads
\begin{equation}\label{eq:cloverWL}
\raisebox{-0.8cm}{\includegraphics[width=2.cm]{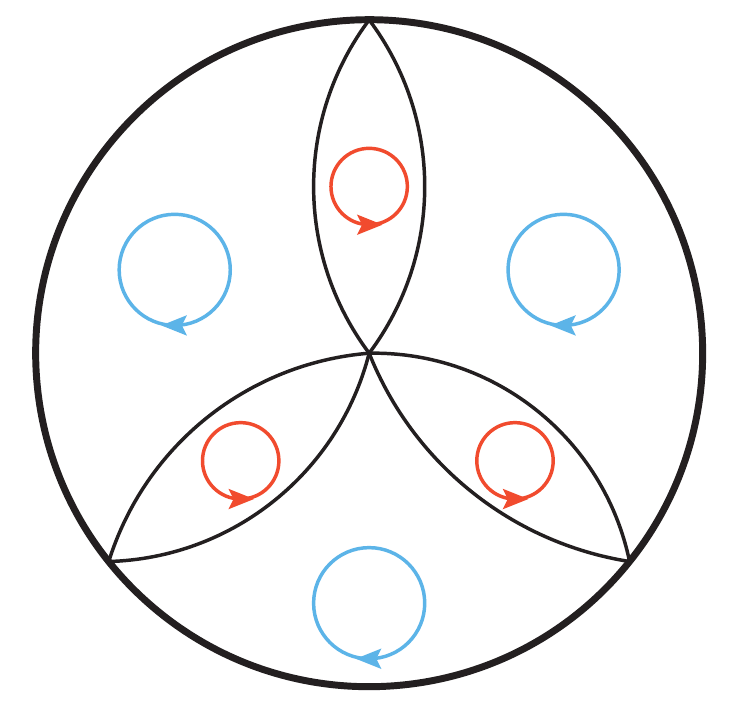}} = -\frac{3\,N_2^2\, \pi^2}{16\, k^2}
\times \raisebox{-0.8cm}{\includegraphics[width=2.cm]{triangle}}
\end{equation}

As mentioned above, the flavor structure of the sextic potential causes a different overall color factor for this diagram, compared to the other contributions. In equations this mismatch amounts to
\begin{align}\label{eq:3ptlimit}
& \frac{N_1\, 8\pi^3}{k^3} \int\displaylimits_{0}^{2\pi}\!\!\int\displaylimits_{0}^{\tau_1}\!\!\int\displaylimits_{0}^{\tau_2} 
 \left\langle \Tr(Y_1\bar Y_2)(\tau_1) \Tr(Y_2\bar Y_3)(\tau_2) \Tr(Y_3\bar Y_1)(\tau_3)  \right\rangle^{(2)}\times
 \nonumber\\& ~~~~~~~~
 \Tr\left( M(\tau_1)M(\tau_2)M(\tau_3) \right)\,\bigg|_{\text{color limit}}\hspace{-12mm} d\tau_1 d\tau_2 d\tau_3 =\raisebox{-0.8cm}{\includegraphics[width=2.cm]{triangleColor2}} - \frac{8}{3}\, \raisebox{-0.8cm}{\includegraphics[width=2.cm]{cloverColor2}}
\end{align}
which is the sought-after expression of the two-loop correction to the structure constant in terms of Wilson loop diagrams, in the color limit.
In this formula, the explicit clover diagram contribution on the right-hand-side, accounts for the total difference between inserting the three-point function in the Wilson loops and the actual latitude expectation value, in the color limit, after multiplying the former by a suitable overall factor.

Upon using \eqref{eq:3ptstruc} to rewrite the left-hand-side suitably and \eqref{eq:trifromMM2} and \eqref{eq:cloverWL} for the expressions in the right-hand-side, equation \eqref{eq:3ptlimit} becomes
\begin{equation}
\frac{\tilde{\cal C}^{(2)}}{\tilde{\cal C}^{(0)}}  \times \raisebox{-0.8cm}{\includegraphics[width=2.cm]{triangle}}\,\,\,
\coleq\,\,\, -\frac{2\pi^2}{3}\, N_2^2\, \times \raisebox{-0.8cm}{\includegraphics[width=2.cm]{triangle}}
+\frac{\pi^2}{2}\,N_2^2 \times \raisebox{-0.8cm}{\includegraphics[width=2.cm]{triangle}}
\end{equation}
from which I extract
\begin{equation}
\frac{\tilde{\cal C}^{(2)}}{\tilde{\cal C}^{(0)}} \coleq -\frac{\pi^2}{6}\,N_2^2
\end{equation}
This is still just the sum of the relevant diagrams contributing to the three-point function, that is the bare one.
Finally, I normalize it with the expressions \eqref{eq:leading} for the two-point functions of the relevant operators and obtain the prediction
\begin{equation}\label{eq:prediction}
\frac{{\cal C}^{(2)}}{{\cal C}^{(0)}} \coleq \frac{\pi^2}{12}\, N_2^2
\end{equation}

\subsection{An aside: color limit of an extremal three-point function}\label{sec:extremal}

\begin{figure}
\centering
\includegraphics[width=6.5cm]{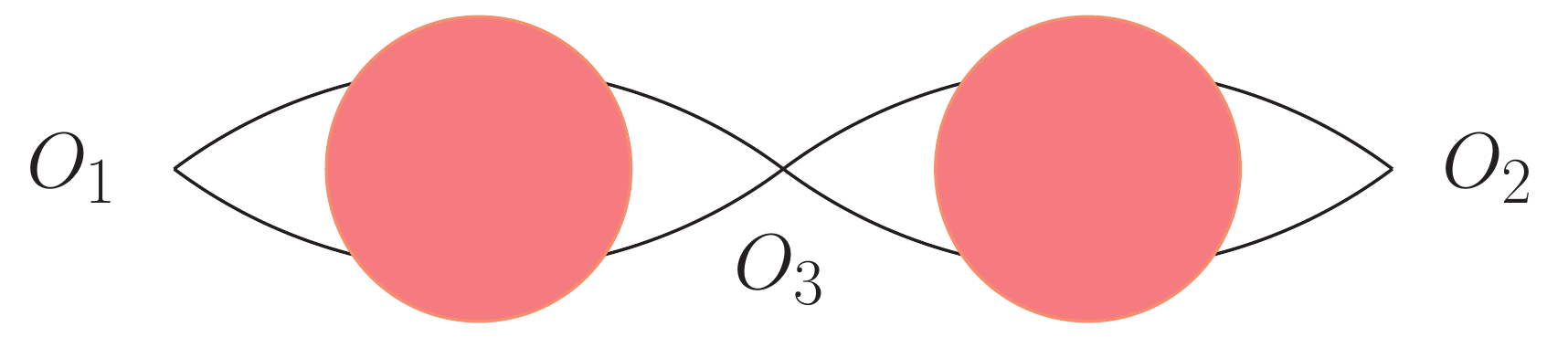}
\caption{Cartoon of the extremal three-point function and its color limit.}\label{fig:extremal}
\end{figure}

In \cite{Young:2014sia} the perturbative computation of an extremal three-point function was performed, up to two loops at weak coupling.
Such a three-point function can be constructed for instance with the twist-one operators $O_1 = \Tr(Y^1 \bar Y_2)$, $O_2 = \Tr(Y^3 \bar Y_4)$ and the length-four protected operator
\begin{equation}
O_3 = \Tr(Y^2 \bar Y_3 Y^4 \bar Y_1) + \Tr(Y^2 \bar Y_1 Y^4 \bar Y_3)
\end{equation}
The corresponding three-point function exhibits the conformal structure
\begin{equation}
\left\langle O_1(x_1) \, O_2(x_2) \,  O_3(x_3) \right\rangle = \frac{\tilde{\cal C}_{\text{extr}}}{|x_{23}|^2\, |x_{13}|^2}
\end{equation}
The structure constant was determined at two-loops in \cite{Young:2014sia}. Here I extend the computation to all orders in the color limit $N_2\gg N_1$, where only powers $N_1 N_2^{2+l}$ at loop $l$ are retained, in the whole structure constant, inclusive of the tree-level factor.
In particular, at tree level there are two possible contractions. One of them is proportional to the color factor $N_1 N_2^2$ and is the relevant one in the color limit. Hence I shall focus on corrections to this structure with leading powers of $N_2$.
By analogous color arguments as above, the corresponding diagrams can only possess the structure of a double leading $N_2$ correction to the scalar tree-level bubbles, illustrated in Figure \ref{fig:extremal}. This is no longer true for the other tree-level color component, but I ignore it here, as subleading in $N_2$.
Using the effective result \eqref{eq:leading}, the un-normalized extremal structure constant evaluates
\begin{equation}
\frac{\tilde{\cal C}_{\text{extr}}}{\tilde{\cal C}_{\text{extr}}^{(0)}} \coleq \left(\frac{\sin \frac{\pi N_2}{k}}{\frac{\pi N_2}{k}}\right)^2
\end{equation}
To normalize it, I divide by the two-point functions of the relevant operators. For length-two chiral primaries this is again given by \eqref{eq:leading}. For the length-4 operator the color limit imposes that the two-point function factorizes into the product of two-independent bi-scalar two-point functions of the same form \eqref{eq:leading}.

Altogether, the normalizations completely cancel the three-point function which in this limit is thus not corrected beyond tree level
\begin{equation}
\frac{{\cal C}_{\text{extr}}}{{\cal C}_{\text{extr}}^{(0)}} \coleq 1
\end{equation}
This is in disagreement with the two-loop result of \cite{Young:2014sia}.
I think I understand where the discrepancy originates. In \cite{Young:2014sia} it is stated that the independent corrections to the bubbles as in Figure \ref{fig:extremal} are exactly cancelled by the normalizations of the length-two operators and hence discarded. I disagree with such a statement and claim instead that the normalizations remove only half of these corrections (operators are normalized by the root of their two-point function to guarantee orthonormality). Indeed, in the color limit I am enforcing, the mismatch between my result and that of \cite{Young:2014sia} can be precisely ascribed to such a different counting of normalizations.
By a similar counting of factors, this argument can be extended to other extremal normalized correlators as well, which are then protected from quantum corrections in the color limit.

\section{Attempted perturbative check}\label{sec:method}

\subsection{Naive attempt}

In this section I attempt a perturbative test of the prediction \eqref{eq:prediction}, which is derived assuming the conjectural matrix model description of the latitude Wilson loop.
At the corresponding lowest non-trivial order, namely two loops, an explicit result for the three-point function exists in literature \cite{Young:2014lka}.
Extracting the relevant color component, such a result shows disagreement with my proposal. In particular it contains a rational term which is absent in \eqref{eq:prediction} and also the part proportional to $\pi^2$ differs.

Given such a disagreement, I first of all performed again the same computation of \cite{Young:2014lka}.
The outcome of such an analysis differs from \cite{Young:2014lka}.
Unfortunately, there are not enough details in \cite{Young:2014lka} to allow me to trace back exactly were our computations depart. I could only nail the discrepancy down to two classes of diagrams for which we have different answers, but I could not understand its precise source any deeper. I was also not able to retrieve the final result of \cite{Young:2014lka} from some intermediate steps that are presented there, so some of the disagreement could just be due to typos. Anyway, the final answer that I got reproducing this computation does not agree with that reported in \cite{Young:2014lka}. 

I do not quote the result of such a computation here, because, as I will explain momentarily, I believe it is not complete and I do not want to clutter this note with incorrect statements.
In particluar, the estimate I obtained following the same computation of \cite{Young:2014lka} does not reproduce \eqref{eq:prediction} either.
This raised my suspicion on the method itself that is used for performing such a computation.
In the rest of this section I aim to clarify such a suspicion, analyze the source of possible issues and devise an alternative strategy to overcome difficulties.

\subsection{The method and its potential issues}

The method for deriving structure constants employed in \cite{Young:2014lka} (and previously adopted in the context of ${\cal N}=4$ SYM in \cite{Plefka:2012rd}) consists in performing a space-time integration on the insertion point of one of the operators.
The underlying idea is that such an operation is easily performed at the level of the general form of the three-point function dictated by conformal symmetry, e.g.~\eqref{eq:3ptstruc} in the case considered here.

When implemented directly on the relevant Feynman diagrams for the computation, the additional integration introduces a dramatic simplification, which makes the computation much simpler and feasible.
In particular, it turns three-point integrals into propagator-type ones, which can be solved efficiently.
The structure constant is then obtained by comparison between the two aforementioned sides of the computation.

I think that such a method is astute and in certain circumstances it proves powerful and effective. I have myself used it heavily to derive results for three-point functions of twist-two operators in ${\cal N}=4$ SYM \cite{Bianchi:2018zal,Bianchi:2019jpy}.
However, as I have already highlighted in \cite{Bianchi:2018zal}, I also believe that there are subtleties connected to the integration process that might put the construction in jeopardy.
The crucial point, which I believe is extremely delicate is as follows and is connected to regularization:
on the side with the general form of the three-point function one integrates a result which is strictly in integer dimensions ($d=3$ in this case) with a $d=3-2\epsilon$ dimensional measure (one can try to extend for example \eqref{eq:3ptstruc} to $d=3-2\epsilon$ analytically continuing the powers of the distances, which works at tree-level, but does not in general at higher perturbative orders).
This is done for two reasons: first, on the side of the general conformal form of the three-point function, the additional integration might be itself divergent and need regularization. This is not the case for the three-point function \eqref{eq:3ptstruc} in three dimensions, but might happen in other settings \cite{Bianchi:2018zal,Bianchi:2019jpy}. Secondly, when performing the integration on the individual Feynman diagrams, the latter are evaluated in dimensional regularization, because they contain divergences, individually. One could try to use a different regulator, however, much of the technical simplifications entailed by this procedure stems from performing integration-by-parts (IBP) reductions and evaluating higher loop momentum integrals, which is much simpler or at least better studied within dimensional regularization.

In particular, the simplicity of the integration method arises when the integration over the insertion point is performed immediately on the Feynman diagram, prior to the other loop integrations.
By doing this, an order-of-limits issue arises.
On one side, one is integrating in $d=3-2\epsilon$ a strictly $d=3$ three-point function, namely an $\epsilon\to 0$ limit has been taken before performing the final integration on the insertion point.
At the level of the Feynman diagrams expansion, instead, the integration over the insertion point is performed first, and the $\epsilon\to 0$ limit is performed at the very end.
Clearly, it is not guaranteed that the two procedures commute.

In previous cases analyzed in ${\cal N}=4$ SYM there are several indications that the method succeeds. In particular they reproduce some results known independently by other methods and satisfy some internal consistency checks, that hint at their correctness.
However there are certain pathological cases where I believe the method may loose predictivity.
In particular, I reckon that the computation of the three-point function \eqref{eq:3ptstruc} in three dimensions falls into such a category. 

\subsection{An example: the clover diagram}\label{sec:clover}

Let me illustrate this point with a clarifying example, that in addition contains some important information that I will use later.

I consider the following peculiar diagram stemming from the sextic scalar potential of ABJM, and which contributes to the three-point function \eqref{eq:3ptstruc}. 
\begin{figure}
\centering
\includegraphics[width=3.5cm]{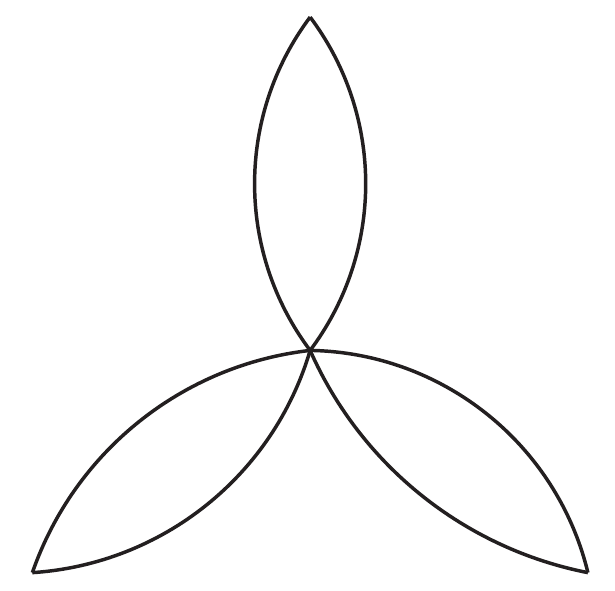}
\caption{Clover diagram constructed out of the sextic scalar vertex of ABJM.}\label{fig:cloverdiag}
\end{figure}
Such a diagram is finite and can be evaluated straightforwardly in $d=3$ dimensions by computing the triangle integral with the method of uniqueness \cite{Kazakov:1984km}.
The result is proportional to the integral
\begin{equation}\label{eq:cloverint}
\int d^3x_0\, \frac{1}{x_{01}^2\, x_{02}^2\, x_{03}^2} = \frac{\pi^3}{(x_{12}^2)^\frac12\, (x_{13}^2)^\frac12\, (x_{23}^2)^\frac12}
\end{equation}
This produces a term possessing already the space-time structure dictated by conformal symmetry, therefore it will contribute to the three-point function \eqref{eq:3ptstruc} non-trivially.
In the present case, the complete result of the diagram reads
\begin{equation}\label{eq:clover}
\raisebox{-0.8cm}{\includegraphics[width=2.cm]{cloverdiagb}} / \,\,\tilde{\cal C}^{(0)} = \frac{\pi ^2 \left(5 N_1^2-12 N_1 N_2 + 5 N_2^2+2\right)}{16\, k^2\, (x_{12}^2)^\frac12\, (x_{13}^2)^\frac12\, (x_{23}^2)^\frac12}
\end{equation}
The right hand side can also be subsequently integrated over one of the insertion points, after re-instating the space-time dependence \eqref{eq:cloverint}, to produce a finite answer.
However, if one takes the same diagram as part of the perturbative expansion of the three-point function \eqref{eq:3ptstruc} and moves the extra integration over an external point to act on the diagram first, before the other integrations (this can be performed safely as long as the integrals are all regulated, for instance in dimensional regularization as here), then one obtains a scale-less integral that vanishes in dimensional regularization.

To summarize I have obtained that the application of the method described above to this particular diagram has produced a vanishing result, and is therefore neglecting a non-trivial contribution \eqref{eq:clover} to the structure constant.
At the technical level it is clear where this discrepancy originates from and is precisely the order-of-limits issue described above.
Moreover, the example points out a general mechanism that can cause problems, which is the presence of sub-integrals (notably bubbles at this perturbative order) that after the additional integration (i.e.~a soft limit in momentum space) become scale-less.

A cheap objection to this reasoning could be that it has been applied to a single diagram, but the final answer when summing all contributions would somehow re-adjust itself to produce the correct result and restore the contribution that is missing.
In fact, the presence of bubbles which are set to zero when applying the method, was already noted in \cite{Young:2014lka} and attributed to a coincidental cancellation of infrared and ultraviolet divergences, which should eventually be innocuous and produce the correct result. 

I do not think this is necessarily the case in general, as I show now.
Imagine modifying the coupling of the sextic vertex in ABJM to an arbitrary value of order $k^{-2}$. This would spoil superconformal symmetry, however, up to two-loop order, the correlation function would still be conformal and satisfy \eqref{eq:3ptstruc} (corrections to the sextic coupling would be higher order). Indeed, as shown above, the contribution originating from this diagram is itself conformal, independently of the rest. Such a contribution would be completely neglected using the integration over an insertion point method, but there is no other diagram that could compensate it, along with the new coupling dependence I have introduced artificially.
This example shows that in fact a physically relevant term can be entirely missed by the method.

Another way of stating the same concept could be considering the theory of a scalar field $\phi$ in three dimensions interacting with a sextic potential $V\sim \phi^6$. The first quantum correction to the three-point functions of operators of the form $\phi^2$ comes from such a vertex and hence is proportional to the structure \eqref{eq:cloverint} and in particular is not vanishing and can be computed without needing to resort to fancy methods. However, performing the same computation using the method of integrating over an insertion point, in dimensional regularization along the lines described above (the space-time structure of this correction is again \eqref{eq:cloverint}, as in the conformal case, so the same prescription would apply), would yield a vanishing result, inconsistently.
If still not convinced, I am providing a further example later on, which will hopefully be even more compelling.

\section{Extension to an operator with spin}\label{sec:twistone}

\subsection{Twist-one operators with spin}

A strong consistency test to ascertain whether the integration-point method is successful, consists in cross checking the extracted structure constants against two inequivalent integrations. This for instance provides compelling checks in ${\cal N}=4$ SYM \cite{Bianchi:2018zal,Bianchi:2019jpy}.
Unfortunately, the three-point function \eqref{eq:3ptstruc} is completely symmetric under the exchange of any operators or $(x_1,x_2,x_3)$ permutations. This means that there are no inequivalent integrations on the external points.

In order to provide such a possibility, I replace one of the operators by a spinning one.
In ABJM twist-one operators can be constructed out of a color singlet pair of scalars and covariant derivatives acting on them.
At a difference with ${\cal N}=4$ SYM these building blocks do not generate a closed sub-sector of operators, rather mixing with fermions occurs. Still, for the specific problem I am considering here, such a mixing does not play an important role and can be mostly neglected. As spelled out below, such a mixing would only kick in for higher order corrections to the three-point functions.

I shall therefore consider two-loop two- and three-point functions involving the operators
\begin{equation}\label{eq:twist1}
\hat O_3^j = \sum_{k=0}^j\, a_{jk}^{\frac{d-3}{2}}\, \Tr\left( \hat D^{j-k} Y^3 \hat D^{k} \bar Y_1 \right) \qquad,\qquad \hat D = D_{\mu} z^{\mu}
\end{equation}
where I have contracted Lorentz indices with a null vector $z$ and indicated such a contraction with a hat. I have chosen the flavors so to replace the operator at position 3 in the previous three-point functions \eqref{eq:3ptstruc}.
An orthonormal set of such operators, which diagonalises the two-point function at tree level can be constructed by choosing the coefficients in \eqref{eq:twist1} to be those of the expansion of Gegenbauer polynomials in three dimensions, or rather in $d=3-2\epsilon$, since I work in dimensional regularization
\begin{equation}\label{eq:Gegenbauer}
\sum_{k=0}^j\, a_{jk}^{\frac{d-3}{2}}\, x^{j-k} y^{k} = (x+y)^j\, \frac{C_j^{\frac{d-3}{2}}}{d-3}\left( \frac{x-y}{x+y} \right)
\end{equation}
In such a form the following tree level two- and three-point functions can be evaluated.
The space-time structure is again fixed by conformal symmetry
\begin{equation}\label{eq:2point}
\left\langle \hat {\cal O}_3^j(0) \doublehat{\bar{\mathcal{O}}} _3 ^k(x) 
\right\rangle = C\, \delta^{jk} \frac{\hat I^j}{|x|^{2+2\gamma_j}}
\end{equation}
where $C$ is the normalization, which depends in general on the coupling constant $k$ and $\gamma_j$ is the anomalous dimension of the operators with spin $j$, yielding the full dimension, inclusive of quantum corrections. The tensor structure is encoded by the object
\begin{equation}\label{eq:Itensor}
\hat I \equiv I_{\mu\nu}\, z_1^{\mu}\, z_2^{\nu} \qquad, \qquad I_{\mu\nu} \equiv \eta_{\mu\nu} - 2\, \frac{x_{\mu}x_{\nu}}{x^2}
\end{equation}
where here I have contracted the two operators with two distinct null vectors $z_1$ and $z_2$.
For practical computations it is convenient to set them equal and extract the normalization $C_j$ from the coefficient of the $\hat x^{2j}$ power.
The space-time dependence of the three-point function is also determined by conformal symmetry and reads
\begin{equation}\label{eq:3ptstructure}
\left\langle {\cal O}_{1}(x_1)\, {\cal O}_{2}(x_2)\, \hat {\cal O}_{3}^j(x_3) \right\rangle = \frac{\tilde{\cal C}_{j}\,\hat Y^j}{|x_{12}|^{1-\gamma_j} |x_{23}|^{1+\gamma_j} |x_{13}|^{1+\gamma_j}}
\end{equation}
where I define
\begin{equation}
\hat Y \equiv Y_{\mu}\, z^{\mu}\qquad,\qquad Y^\mu \equiv \frac{x_{23}^\mu}{x_{23}^2} - \frac{x_{13}^\mu}{x_{13}^2}\qquad,\qquad
x_{ij} \equiv x_i - x_j 
\end{equation}
and the structure constant $\tilde{\cal C}_j$ is not normalized by the two-point functions of the operators, as opposed to ${\cal C}_j$.
At weak coupling, I consider the expansions for $k\gg 1$
\begin{equation}
C_j = \sum_{i=0}^{\infty}\, \frac{C_{j}^{(2i)}}{k^{2i}} \qquad,\qquad
{\cal C}_{j} = \sum_{i=0}^{\infty}\, \frac{{\cal C}_{j}^{(2i)}}{k^{2i}}
\end{equation}
in which I am anticipating the vanishing of odd loop order corrections.
At two loops these constants can be further divided into two independent color components
\begin{equation}\label{eq:colornot}
C_j^{(2)} = N_1 N_2 \left(N_1^2 + N_2^2 - 2\right) C_{j,N_1^2}^{(2)} + N_1 N_2 \left(N_1 N_2 - 1 \right) C_{j,N_1N_2}^{(2)}
\end{equation}
and analogously for ${\cal C}$. The overall color coefficient $N_1 N_2$ stems from the tree-level contribution. As it can be seen in \eqref{eq:colornot}, no planar limit is enforced at any point in the computation.

According to the definitions above, the tree level normalization of the two-point function and the structure constant read $(j\geq 1)$
\begin{align}
C_j^{(0)} &= \frac{N_1 N_2}{(4\pi)^2}\, \frac{2^{j}\, \Gamma\left(2j\right)}{j}\\
\tilde{\cal C}_{j}^{(0)} &= \frac{N_1 N_2}{(4\pi)^3}\, \frac{2^j\, \Gamma\left(j+\frac12\right)}{\sqrt{\pi}\, j}
\end{align}
as can be calculated straightforwardly, making use of the identity (for $d\to 3$)
\begin{equation}
a_{jk}^{\frac{d-3}{2}} \to \frac{2\, (-1)^k\, \Gamma(2j)}{\Gamma\left( 2k+1 \right)\, \Gamma\left( 2j-2k+1 \right)}
\end{equation}
In the following, I will consider two-loop corrections to two- and three-point functions and always take their ratios with the tree-level counterparts so that the latter will not play any further role.
Note that both $C_j^{(0)}$ and $\tilde{\cal C}_j^{(0)}$ diverge for $j\to 0$ as a consequence of the normalization \eqref{eq:Gegenbauer}.
Still, the normalized three-point function (that is normalized by the two-point functions of the operators), which is the physical relevant quantity I am after, exhibits a smooth limit for vanishing spin, as it should be and which I will use momentarily.

\subsection{Two-loop correction to the two-point function}

I now compute quantum corrections to these objects up to two loops. One loop is trivial and I will jump to two directly.
I start with the two-point functions \eqref{eq:2point}.
At two loops a variety of diagrams appears. A taste of them could be roughly gotten from Figure \ref{fig:diagrams} (relative to the three-point function), removing the third operator and getting rid of equivalent configurations (though there are some additional graphs not obtainable from those). For chiral primaries I refer to \cite{Young:2014lka} for a more detailed computation, with which I agree. For the case of operators with spins, an additional gauge ladder diagram contributes, which vanishes identically for two chiral primaries. Moreover, contributions from the gauge fields in the covariant derivatives have to be taken into account.
All computations are performed in momentum space, where I make use of integration-by-parts (IBP) identities \cite{Tkachov:1981wb,Laporta:1996mq,Laporta:2001dd} to reduce integrals to a set of master integrals. This is especially useful when dealing with the derivatives implied in the definition of operators with spin \eqref{eq:twist1}, but the complexity of the process rapidly increases with the number of such derivatives, that is the spin. In practice I have used the FIRE6 implementation \cite{Smirnov:2008iw,Smirnov:2019qkx} in conjunction with LiteRed \cite{Lee:2012cn,Lee:2013mka} for carrying out IBP reductions.

The individual relevant Feynman integrals needed for the two-loop computation are in general divergent. They are regulated with dimensional regularization and the dimensional reduction scheme \cite{Siegel:1979wq}, consisting of performing the numerator tensor algebra in strictly three dimensions prior to integrating in $d=3-2\epsilon$. Such a scheme preserves gauge invariance \cite{Chen:1992ee} and has proven successful in a number of perturbative applications in ABJM \cite{Minahan:2009wg,Bianchi:2013rma,Bianchi:2013zda,Griguolo:2013sma,Bianchi:2013pva,Bianchi:2014iia}.

At two-loop order an ultraviolet divergence appears in the two-point functions of the bare spinning operators \eqref{eq:twist1}, since they are not protected.
In particular, they \eqref{eq:twist1} mix under renormalization with operators of the same spin and dimension. These could be conformal descendants constructed applying total derivatives on the operators \eqref{eq:twist1} and operators constructed out of fermions and derivatives applied on them (and their descendants). 
The bare operators \eqref{eq:twist1} have then to be renormalized multiplicatively by certain matrices that encode mixing
\begin{equation}
\hat {\cal O}_j = Z_{jk}\, \hat \partial^{j-k} \hat O_{k} + \text{fermions}
\end{equation}
Such matrices include a finite renormalization in order to achieve two-loop orthogonal two-point functions, as in \eqref{eq:2point}, also at loop level. Determining this finite contribution usually requires imposing diagonalizing higher order divergences.
In this particular case, demanding two-loop orthogonality already fixed the finite two-loop contribution that automatically ensures higher loop diagonal divergences.
These renormalized operators are those relevant for computing conformal three-point functions, which obey the structure \eqref{eq:3ptstructure} and which I aim at computing.
In particular, to the perturbative order I am working (which is the first non-trivial), mixing with fermions does not play a prominent role, as tree-level three-point functions of two chiral primary protected operators \eqref{eq:operators} and a length-two operator constructed with fermions are vanishing since there are no possible Wick contractions. Instead, mixing with descendants of the operators \eqref{eq:twist1}, plays an important role for the following calculations, and has to be taken into account.

Further, the two-point functions \eqref{eq:2point} can be made orthonormal by including a finite renormalization of the operators on the diagonal of the mixing matrix, that sets $C_j=1$. The resulting operators are those providing the proper normalized three-point function, which is the ultimate goal of the computation. Still, in the following it will be convenient to also compute un-normalized structure constants $\tilde{\cal C}_j$. To clarify, by these I mean that the operators have been consistently orthogonalized as in \eqref{eq:2point}, but their finite diagonal normalization has not been removed to 1. The reason why I consider such structure constants as well is that thanks to a technical detail to be explained below, I am able to compute them to higher values of the spin. Moreover, the un-normalized structure constant of three chiral primary operators also played a more fundamental role in the comparison with the matrix model of the previous section. In particular, it is the quantity naturally contributing to the expectation value of the latitude Wilson loop.

The anomalous dimension of the operators \eqref{eq:twist1} can be extracted from the bare two-point functions and evaluates for spin $j$
\begin{equation}\label{eq:dimensions}
\gamma_j =  \frac{4(N_1N_2-1)}{k^2}\left( S_1(j) - S_{-1}(j) - \frac{1-(-1)^j}{2\, j}  \right) + O\left(k^{-4}\right)
\end{equation}
As a check, I have verified that the color leading piece coincides with that expected from the two-loop dilatation operator \cite{Zwiebel:2009vb,Minahan:2009te}, which in turn cuncurs with the prediction from the Bethe ansatz of \cite{Minahan:2008hf,Gromov:2008qe}.
I have not found in literature an explicit solution for the anomalous dimensions of such operators for generic spins as in \eqref{eq:dimensions}. I derived it here, after performing a few checks at finite values of the spin, from which the pattern \eqref{eq:dimensions} can be inferred.
This constitutes a successful partial check of my computation.

Next, in table \ref{tab:2pt} I spell out the finite diagonal part of the renormalization matrix $Z_{jj}$ for the two-point functions of the operators \eqref{eq:twist1}, which I have computed up to spin 6.
\begin{table}
\begin{tabular}{c|ccccccc}
$j$ & 0 & 1 & 2 & 3 & 4 & 5 & 6 \\[2pt]
\hline\\[-2ex]
$Z_{jj, N_1^2}^{(2)}\big|_{\text{finite}}$ & $\displaystyle\frac{\pi ^2}{12} $ & $\displaystyle\frac{\pi ^2}{12} $ & $\displaystyle\frac{1}{12}+\frac{\pi ^2}{12} $ & $\displaystyle\frac{1}{6}+\frac{\pi ^2}{12} $ & $\displaystyle\frac{407}{1680}+\frac{\pi ^2}{12} $ & $\displaystyle\frac{521}{1680}+\frac{\pi ^2}{12} $ & $\displaystyle\frac{15439}{41580}+\frac{\pi ^2}{12}$ \\[12pt]
$Z_{jj, N_1 N_2}^{(2)}\big|_{\text{finite}}$ & $0 $ & $1 $ & $0 $ & $\displaystyle -\frac{8}{9} $ & $\displaystyle -\frac{445}{252} $ & $\displaystyle -\frac{6847}{2700} $ & $\displaystyle -\frac{7179}{2200}$ 
\end{tabular}\caption{Finite diagonal part of the two-loop renormalization matrix up to spin 6. This will be used to normalize the structure constants. The color notation is the same as in \eqref{eq:colornot}}\label{tab:2pt}
\end{table}
These will be used for normalizing the structure constants.

\subsection{Two-loop correction to the structure constant}

Next, I compute the two-loop correction to the three-point function \eqref{eq:3ptstructure} using the method of integration over an external point.
The relevant diagrams are illustrated in Figure \ref{fig:diagrams}, which have been generated and computed automatically.
\afterpage{\begin{figure}[t]
\centering
\includegraphics[width=\textwidth]{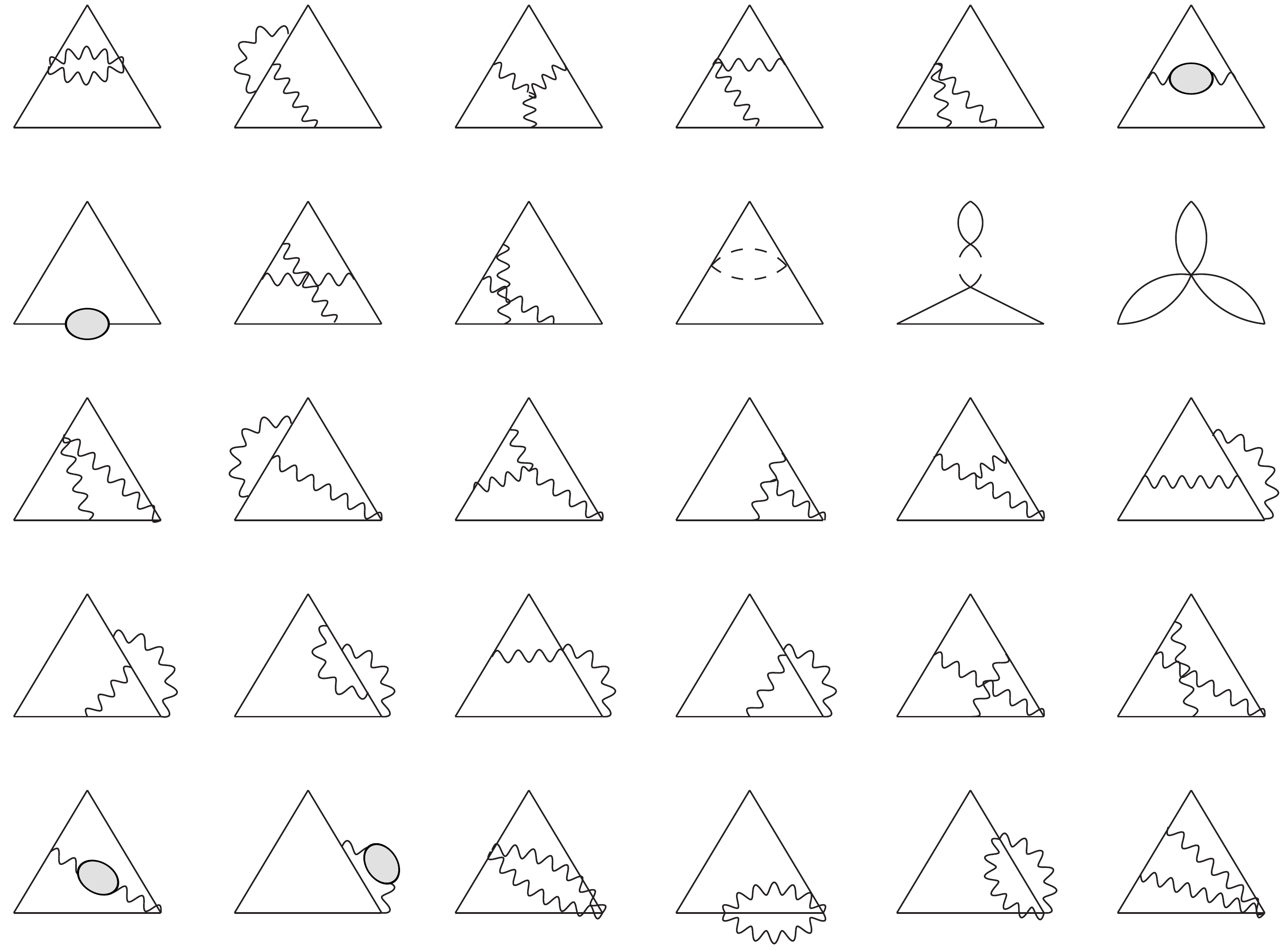}
\caption{Diagrams contributing to the three-point function of two chiral primary twist-one operators and one with spin at position 3, which is the lower-right corner of the triangles. Solid, dashed and wavy lines represent scalars, fermions and gluons, respectively.
Only non-vanishing diagrams are shown, though some of them do vanish when applying the integration method on one of the external insertion points.
Blobs represent self-energy corrections, two-loop for the scalar field and one-loop for the gauge propagator. The complete expression for the former can be found in \cite{Bianchi:2017afp,Bianchi:2017ujp}.
Part of the diagrams emerge from the gauge field entering the covariant derivatives applied to the operator with spin. These diagrams are not present in the analogous computation for three chiral primary operators. 
}\label{fig:diagrams}
\end{figure}
}
Only a representative diagram for each contraction has been shown, apart from one contribution that I repeated just to fit all graphs into a nice grid. Permutations thereof have to be included and these can be inequivalent, since in the presence of an operator with spin, the diagrams only possess a $\mathbb{Z}_2$ symmetry. Further, when applying the integration method, the vertex of the corresponding operator is replaced by a doubled propagator or by an effective vertex, depending on the connectivity of the original operator (whether it was 2, or higher). This introduces an additional asymmetry, basically making most of permutations inequivalent and producing a large number of different contributions, from the relatively small set of diagrams shown above.

As a consistency test, all computations have been checked to be gauge invariant. Moreover, the divergent piece of the bare three-point function allows for an independent derivation of the anomalous dimensions of twist-one operators, which coincides with that extracted from the two-point function \eqref{eq:dimensions}. 

\paragraph{Inequivalent integrations}
In the case with an un-protected operator with spin, two inequivalent integrations are possible, namely over the operator with spin, or over a protected one.
At the level of integrating the general structure of the three-point function \eqref{eq:3ptstructure}, the two integrations produce respectively
\begin{align}
&\int d^{3-2\epsilon}x_2\, \frac{\tilde{\cal C}_{j}}{|x_{12}|^{1-\gamma_j} |x_{23}|^{1+\gamma_j} |x_{13}|^{1+\gamma_j}} \left( \frac{\hat x_{13}}{x_{13}^2} - \frac{\hat x_{23}}{x_{23}^2} \right)^j
= -\frac{\pi^{\frac32}\, \Gamma(j+1)\, \hat x_{13}^j}{\Gamma\left(j+\frac12\right)\, (x_{13}^2)^j}\times\label{eq:int1}
\\& \left(2\tilde{\cal C}_j^{(0)} 
- \frac{1}{k^2} \left(
\gamma_j^{(2)}\, \tilde{\cal C}_j^{(0)}\, \left(\psi\left(j+\frac{1}{2}\right)-\psi(j+1) +\log 4\right)-2\tilde{\cal C}_j^{(2)}\right)\right) + O(\epsilon) + O(k^{-4})
\end{align}
and
\begin{align}
&\int d^{3-2\epsilon}x_3\, \frac{\tilde{\cal C}_{j}}{|x_{12}|^{1-\gamma_j} |x_{23}|^{1+\gamma_j} |x_{13}|^{1+\gamma_j}} \left( \frac{\hat x_{13}}{x_{13}^2} - \frac{\hat x_{23}}{x_{23}^2} \right)^j
= \frac{ 4^j\, \pi\, \hat x_{12}^j }{(2 j-1)\, (x_{12}^2)^j}\times\label{eq:int2}\\&
\left(\tilde{\cal C}_j^{(0)} 
+ \frac{1}{k^2} \left(
\gamma_j^{(2)}\, \tilde{\cal C}_j^{(0)}\, \left(\psi\left(j-\frac{1}{2}\right)-\psi(2 j) + 1+\log 4\right)+\tilde{\cal C}_j^{(2)}\right)\right) + O(\epsilon) + O(k^{-4})\nonumber
\end{align}
where $\gamma_j^{(2)}$ stands for the anomalous dimension at two loops, that is basically \eqref{eq:dimensions}, and $\psi$ is the digamma function.
The first expression is well-defined for $j>0$, for $j=0$ both right-hand-sides  collapse to $-2\pi\, \tilde{\cal C}$.
These different integrations generate two distinct perturbative expansions at the level of the individual Feynman diagrams, which should eventually yield the same structure constants, upon comparing with \eqref{eq:int1} and \eqref{eq:int2}, respectively, if the method is working properly.
In particular, integrating over the spinning operator insertion point, makes mixing with descendants negligible, as in momentum space it corresponds to a soft limit, which trivializes them all.
On the contrary, mixing has to be kept into account when integrating over the insertion point of a protected operator.
This technical difference makes the computation using the first integration easier. In particular, since there is no need to know the non-diagonal part of the mixing matrix, the operators can be renormalized directly by their anomalous dimension \eqref{eq:dimensions} and the un-normalized structure constant is obtained. 

In practice, the bottleneck of the computations arises in the IBP reduction, whose complexity grows rapidly with the number of derivatives. This means that the un-normalized structure constants at spin $j$, which entails $j$ additional derivatives, has roughly the same level of complexity as that of a two-point function of operators of spin $j/2$, which involves $j$ derivatives too. Consequently, the former computation can be pushed to higher values of the spin compared to the latter, at fixed IBP reduction complexity. 

\paragraph{A discrepancy}
Computing the first few spins explicitly, by integrating over the insertion point of the operator with spin, I derived the following results for the un-normalized two-loop structure constants of \eqref{eq:3ptstructure} (divided by the tree level contribution), which are reported in Table \ref{tab:3pt}.
\begin{table}
\begin{tabular}{c|ccccc}
$j$ & 1 & 2 & 3 & 4 & 5 \\[2pt] \hline \\[-12pt]
$\frac{\tilde{\cal C}_{j,N_1^2}^{(2)}}{\tilde{\cal C}_j^{(0)}}$ & ${\color{red}\frac{\pi^2}{4}}-\frac{5 \pi ^2}{12}$ & ${\color{red}\frac{\pi^2}{4}}-\frac{1}{12}-\frac{5 \pi ^2}{12}$ & ${\color{red}\frac{\pi^2}{4}}-\frac{1}{6}-\frac{5 \pi ^2}{12}$ & ${\color{red}\frac{\pi^2}{4}}-\frac{407}{1680}-\frac{5 \pi ^2}{12}$ & ${\color{red}\frac{\pi^2}{4}}-\frac{521}{1680}-\frac{5 \pi ^2}{12}$  \\[10pt]
$\frac{\tilde{\cal C}_{j,N_1 N_2}^{(2)}}{\tilde{\cal C}_j^{(0)}}$ & $-2+\frac{\pi ^2}{2}$ & $-\frac{5}{3}+\frac{\pi ^2}{2}$ & $-\frac{71}{90}+\frac{\pi ^2}{2}$ & $\frac{7}{180}+\frac{\pi ^2}{2}$ & $\frac{3842}{4725}+\frac{\pi ^2}{2}$   \\[15pt]
\hline\hline
\end{tabular}

\begin{tabular}{c|ccc}
$j$ & 6 & 7 & 8  \\[2pt] \hline \\[-12pt]
$\frac{\tilde{\cal C}_{j,N1^2}^{(2)}}{\tilde{\cal C}_j^{(0)}}$ & $-\frac{15439}{41580}-\frac{5 \pi ^2}{12}{\color{red}+\frac{\pi^2}{4}}$ & $-\frac{35501}{83160}-\frac{5 \pi ^2}{12}{\color{red}+\frac{\pi^2}{4}}$ & $-\frac{826445}{1729728}-\frac{5 \pi ^2}{12}{\color{red}+\frac{\pi^2}{4}}$   \\[10pt]
$\frac{\tilde{\cal C}_{j,N_1 N_2}^{(2)}}{\tilde{\cal C}_j^{(0)}}$ & $\frac{636907}{415800}+\frac{\pi ^2}{2}$ & $\frac{82991107}{37837800}+\frac{\pi ^2}{2}$ & $\frac{212909303}{75675600}+\frac{\pi ^2}{2}$   \\[15pt]
\hline\hline\\[-12pt]
$j$ & 9 & 10 & 11  \\[2pt] \hline \\[-12pt]
$\frac{\tilde{\cal C}_{j,N1^2}^{(2)}}{\tilde{\cal C}_j^{(0)}}$ & $-\frac{4537901}{8648640}-\frac{5 \pi ^2}{12}{\color{red}+\frac{\pi^2}{4}}$ &  $-\frac{7936296197}{13967553600}-\frac{5 \pi ^2}{12}{\color{red}+\frac{\pi^2}{4}}$ & $-\frac{8502782417}{13967553600}-\frac{5 \pi ^2}{12}{\color{red}+\frac{\pi^2}{4}}$  \\[10pt]
$\frac{\tilde{\cal C}_{j,N_1 N_2}^{(2)}}{\tilde{\cal C}_j^{(0)}}$ & $\frac{1006235311}{296881200}+\frac{\pi ^2}{2}$ & $\frac{7212113723}{1833241410}+\frac{\pi ^2}{2}$ & $\frac{27578945077}{6204817080}+\frac{\pi ^2}{2}$ \\[15pt]
\hline\hline\\[-12pt]
$j$ & 12 & 13 & 14 \\[2pt] \hline \\[-12pt]
$\frac{\tilde{\cal C}_{j,N1^2}^{(2)}}{\tilde{\cal C}_j^{(0)}}
$ & $-\frac{51942269699}{80313433200}-\frac{5 \pi ^2}{12} {\color{red}+\frac{\pi^2}{4}} $  & $-\frac{13703103311}{20078358300}-\frac{5 \pi ^2}{12}{\color{red}+\frac{\pi^2}{4}}$ & $-\frac{652072423}{910435680}-\frac{5 \pi ^2}{12}{\color{red}+\frac{\pi^2}{4}}$ \\[10pt]
$\frac{\tilde{\cal C}_{j,N_1 N_2}^{(2)}}{\tilde{\cal C}_j^{(0)}}$ & $\frac{7318927163869}{1484192245536}+\frac{\pi ^2}{2}$  & $\frac{13001915568904289}{2411812398996000}+\frac{\pi ^2}{2}$ & $\frac{42191855210537597}{7235437196988000}+\frac{\pi ^2}{2}$
\end{tabular}\caption{Two-loop un-normalized structure constants of \eqref{eq:3ptstructure}, derived from integrating over the insertion point of the operator with spin. The color components are separated for each spin and the ratio is taken by the tree-level contribution. As argued below, these results are missing a constant totally symmetric contribution that is neglected spuriously because of the extra integration. Then, the correct result should be gotten by adding a further $\frac{\pi^2}{4}$ to the color components $\frac{\tilde{\cal C}_{j,N1^2}^{(2)}}{\tilde{\cal C}_j^{(0)}}$ which is included in the table in red. Neglecting this piece, the result from the naive integration over $x_3$ $\frac{\tilde{\cal C}_{j}^{(2)}}{\tilde{\cal C}_j^{(0)}}\,\,\Big|_{\int d^dx_3}$ can be read.}\label{tab:3pt}
\end{table}
When computing the same quantity, but from the integration over the position of a scalar operator, a different result emerges. 
After a few spins, one realizes that the difference (integration over the position of the operator with spin minus that of the scalar) evaluates to a constant
\begin{equation}\label{eq:mismatch}
\frac{\tilde{\cal C}_j^{(2)}}{\tilde{\cal C}_j^{(0)}}\,\,\Big|_{\int d^dx_3} - \frac{\tilde{\cal C}_j^{(2)}}{\tilde{\cal C}_j^{(0)}}\,\,\Big|_{\int d^dx_2} = -\frac{(N_1-N_2)^2\, \pi^2}{4\,k^2}
\end{equation}
This shows that something is going wrong with the method described above. From the example provided in section \ref{sec:clover}, this can be ascribable to the fact that certain contributions can go missing when the integration method is applied and it somehow clashes with an order-of-limits issue and dimensional regularization. The mismatch would not be visible on the ABJM slice $N_1=N_2$, but I regard this as an accidental feature, rather than an effect with some deep meaning.

\paragraph{Analysis of the discrepancy}
It would be natural to consider the clover diagram of Figure \ref{fig:cloverdiag} as a natural candidate for the discrepancy, as I already know it is problematic.
However this cannot be responsible for the mismatch \eqref{eq:mismatch}, as the clover diagram can be shown to vanish identically when one of the operators has spin. In practice, the distribution of derivatives is precisely designed for this to happen (on the single bubble where the spinning operator is inserted), as this is basically the same condition which prevents mixing between an operator with spin and a scalar. Moreover, the color structure of \eqref{eq:mismatch} does not coincide with that of the clover diagram, which means that some additional contribution is required.

Guided by some intuition and the peculiar color structure of \eqref{eq:mismatch}, another natural candidate is the diagram of Figure \ref{fig:tri-bub2}.
\afterpage{\begin{figure}[ht!]
\centering
\includegraphics[width=3.5cm]{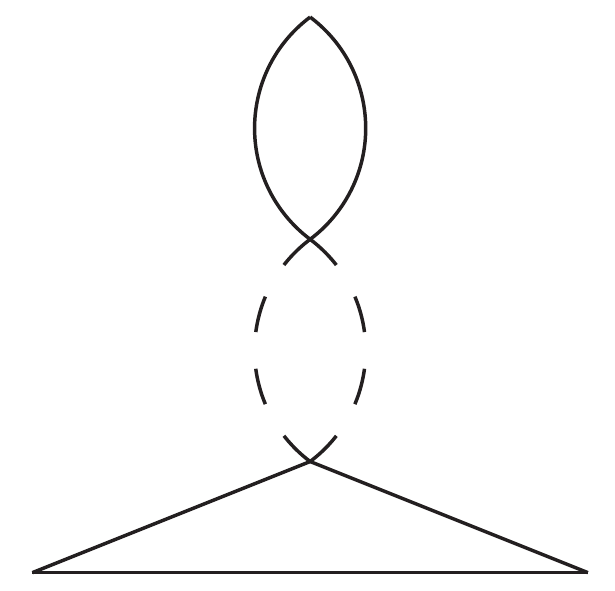}
\caption{Diagram constructed out of two Yukawa vertices that is neglected when integrating over the exterior tip of the bubble.}\label{fig:tri-bub2}
\end{figure}
}
Indeed it contains a bubble where an operator is inserted. Integrating over that operator yields again a scale-less integral which vanishes in dimensional regularization, which can bring problems as in the clover integral case.

In fact, such a diagram is again easy to evaluate individually, without performing the extra integration over an insertion point.
The result is finite and the relevant integrals can again be evaluated with the method of uniqueness. For instance, in the case where the third operator has vanishing spin the integral reads
\begin{align}
\int d^3x_4\,d^3x_5\, \frac{1}{(x_{23}^2)^\frac12\, (x_{24}^2)^\frac12\, (x_{34}^2)^\frac12\, (x_{45}^2)^2\, x_{15}^2} &= 
-\int d^3x_5\, \frac{2\pi}{x_{25}^2\, x_{35}^2\, x_{15}^2} =
\nonumber\\&=
- \frac{2\pi^4}{(x_{12}^2)^\frac12\, (x_{13}^2)^\frac12\, (x_{23}^2)^\frac12}
\end{align}
After including the relevant factors and summing over three equivalent permutations, this diagram in the $j=0$ case produces the final result
\begin{equation}
\raisebox{-0.8cm}{\includegraphics[width=2.cm]{tri-bub2b}}\,\, /\,\,\, \tilde{\cal C}^{(0)} + \text{permutations} = -\frac{(N_1-N_2)^2\, \pi^2}{4\,k^2\, (x_{12}^2)^\frac12\, (x_{13}^2)^\frac12\, (x_{23}^2)^\frac12} \times 3 \qquad\qquad j=0
\end{equation}
Extending the computation to include an operator with spin, the addition of derivatives makes the three permutations inequivalent.
For the same explanation as above, when a bubble ends on such an operator with spin the diagram vanishes identically, therefore only two equivalent permutations are non-vanishing, when in the presence of a spinning operator.
The extra derivatives acting on certain pairs of propagators in \eqref{eq:twist1} can be moved outside of the integrals and applied later on the integrated quantity, in such a way that they re-construct a tree-level contribution.
Hence, taking the ratio with the tree-level correlator and summing the two non-vanishing permutations yields the total contribution
\begin{equation}
\raisebox{-0.8cm}{\includegraphics[width=2.cm]{tri-bub2b}}\,\, /\,\,\, \tilde{\cal C}_j^{(0)} + \text{permutations} = -\frac{(N_1-N_2)^2\, \pi^2}{4\,k^2\, (x_{12}^2)^\frac12\, (x_{13}^2)^\frac12\, (x_{23}^2)^\frac12} \times 2 \qquad\qquad j>0
\end{equation}
When applying the integration on insertion point method on this diagram there is an additional effect, as mentioned above: one of the three permutations of the diagrams is forced to vanish, namely that where the bubble connects to the integrated point.
In the $j=0$ case, this effectively removes one of the three permutations, which is therefore overlooked by this procedure.
For non-vanishing spin, when the integration is performed over the insertion point of an operator with spin this integration is harmless, as the corresponding diagram already vanishes identically, for the reason explained above. Therefore, no non-vanishing contribution is overlooked by the integration method. On the contrary, when integrating over a protected operator, the corresponding diagram is forced to zero, spuriously.

The net effect of such a discrepancy is that the evaluation of the diagram of Figure \ref{fig:tri-bub2} is performed incorrectly when integrating over $x_2$ and differs from the other by
\begin{equation}\label{eq:tribub}
\raisebox{-0.8cm}{\includegraphics[width=2.cm]{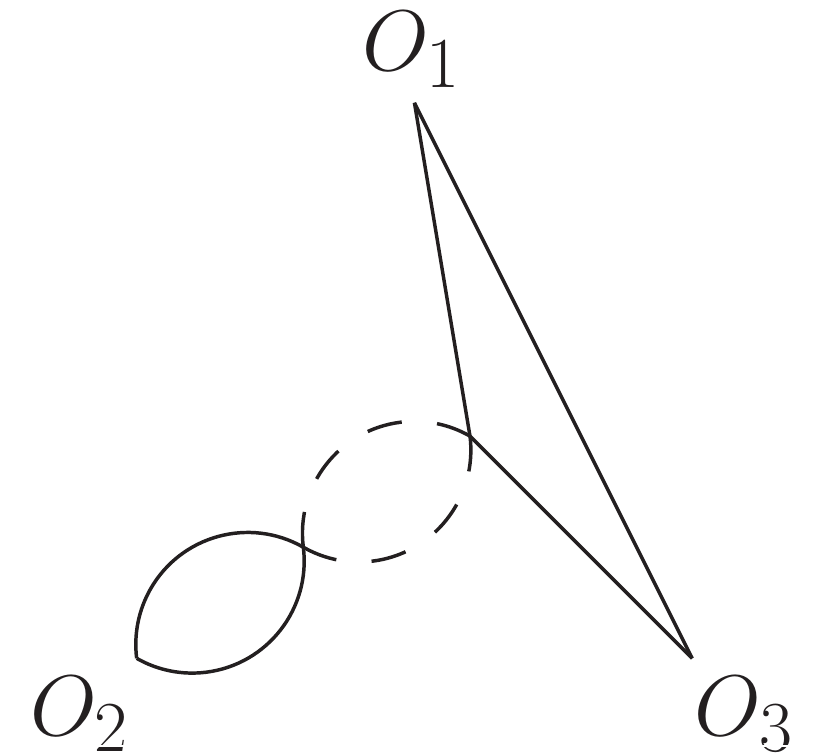}}\,\, /\,\,\, \tilde{\cal C}_j^{(0)} = -\frac{(N_1-N_2)^2\, \pi^2}{4\,k^2\, (x_{12}^2)^\frac12\, (x_{13}^2)^\frac12\, (x_{23}^2)^\frac12}
\end{equation}
This precisely coincides with the whole difference between structure constants pointed out in \eqref{eq:mismatch}, that is diagram \ref{fig:tri-bub2} is likely to encompass the whole mismatch introduced by the method.
In particular, if there were any other similar mismatches caused by the order-of-limits issue at any other point of the calculation (and that are not completely symmetric over the exchange of the operators), the comparison between the two integration points would have detected them.

\paragraph{Further sources of discrepancy}
Still, I cannot completely rule out the possibility of a further correction which is overlooked evenly by the two different integrations. This can be for instance of the form of the clover diagram which is completely symmetric. 
Actually, at two loops, this is the only integral with bubbles that can become tadpoles after the integration on an external point and which is completely symmetric, therefore any additional missing terms can only come from such integrals.
Even though I have argued before that the contribution of the sextic potential diagram does not apply when one of the operators possesses spin, that argument referred to the clover Feynman diagram itself, however the same integral structure might emerge from other diagrams when performing their algebra, for instance when some propagators are cancelled by numerator factors, or even at the level of IBP reductions of derivatives.

I will for the moment assume that the result derived from integrating over the insertion point of the operator with spin is correct up to such clover-like terms and use it to derive the corresponding part of the spin 0 result as a limit. I recall that it was not possible to perform such a check for the $j=0$ case, as all integrations over the insertion points are equivalent and cannot detect any issue introduced by the method.
Finally, I will try to fix the additional pieces involving clover-like integrals and add them by hand to obtain the whole structure constant.

\subsection{The final result for the structure constant}

\paragraph{The vanishing spin limit}
Taking the limit for vanishing spin of the results above is not trivial, as I have not been able to fix the general dependence of the numbers in table \ref{tab:3pt} on the spin $j$.
Presumably it is given by some combination of harmonic sums, but I have not been able to determine it. From analyzing the highest prime factor in the denominator, I infer that such a combination should contain a depth-1 harmonic sum with argument of order $2j$. Furthermore, plotting the data seems to exhibit an oscillatory behavior between odd and even spins, which is no surprise, since this already happens at the level of the anomalous dimensions. 

Still, it seems clear that the terms proportional to $\pi^2$ can be safely considered constant in $j$.
Moreover, the rational terms of the $N_2^2$ color component (first row of table \ref{tab:3ptj}) seem to be in tight relation with those of the corresponding two-point function (first row of table \ref{tab:2pt}), which, incidentally, produces the cancellation of the rational part for this color component of the normalized structure constant.
I will assume the rational part of the $N_2^2$ color component of the structure constant to converge to 0 for spin 0 as well.

For the other color component the limit is not as clear. Performing the same computation as above, integrating on $x_3$, at $j=0$ and adding by hand an additional contribution \eqref{eq:tribub}, which I argued above would have been overlooked by naively applying the integration method, one finds precisely the same coefficient as for the spin $j=1$ structure constant. I will take this as the limit for $j\to 0$.
With these considerations the limit of the structure constant for the case of three chiral primary operators is derived, but according to my reasoning this could still miss contributions from clover type integrals, which could not have been detected by the analysis so far.

\paragraph{Fixing clover-like integrals}
I finally look for clover integrals that have not been taken into account yet, because they are forcefully set to zero by the integration method.
Such contributions will have to be added to the partial results for the structure constants determined above with the insertion point integration method $\frac{\tilde{\cal C}_{0}^{(2)}}{\tilde{\cal C}_0^{(0)}}\,\,\bigg|_{\int d^dx_3}$, by hand.
As derived in section \ref{eq:clover}, there is indeed a contribution from the sextic potential diagram, which contributes at $j=0$, where it evaluates \eqref{eq:clover},  and vanishes for $j>0$.
Further, it can be ascertained that the same integral also emerges after cancelling propagators against numerator factors in diagrams three and four in the first line of Figure \ref{fig:diagrams}.
They read
\begin{align}\label{eq:extraclover}
\raisebox{-0.8cm}{\includegraphics[height=1.75cm]{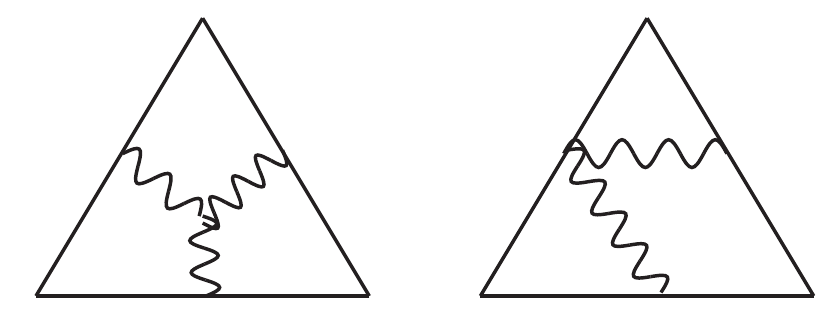}} \bigg|_{\raisebox{-0.3cm}{\includegraphics[width=0.75cm]{cloverdiagb}}\,,\, j=0} &= \tilde{\cal C}_0^{(0)} \Bigg(\frac{N_1^2 + N_2^2 -2}{8\, \pi\, k^2} 
\\&
- \frac{3\left( N_1^2 + N_2^2 - 4N_1 N_2 +2 \right)}{16\, \pi\, k^2} \Bigg)\, \raisebox{-0.8cm}{\includegraphics[width=2.cm]{cloverdiagb}}\nonumber
\end{align}
where the picture on the right-hand-side represent here only the integral \eqref{eq:cloverint}, not the sextic potential diagram. The integral effectively contributes with a factor $\pi^3$, that multiplied by the quantity in the parenthesis, produces the net additional contribution to the ratio of the two-loop structure constant by the tree-level result, which I claim should come from the diagrams \eqref{eq:extraclover}.
Adding these pieces from the diagrams in \eqref{eq:extraclover} only (and not those from the sextic potential graph \eqref{eq:clover} yet) and restricting to the leading $N_2$ color component produces
\begin{equation}
\frac{\tilde{\cal C}_{0}^{(2)}}{\tilde{\cal C}_0^{(0)}}\,\,\bigg|_{\int d^dx_3} + \eqref{eq:extraclover} \coleq 
-N_2^2\frac{23 \pi^2}{48}
\end{equation}
This should correspond, if my conjectural relation to the matrix model is correct, to the scalar triangle diagrams in the color limit of the latitude Wilson loop \eqref{eq:trifromMM}, after subtracting the contribution of the clover diagram in the Wilson loop expectation value, which was computed in \eqref{eq:cloverWL}
\begin{equation}\label{eq:cloverintegral}
\frac{\raisebox{-0.8cm}{\includegraphics[width=2.cm]{triangleColor2}} - \raisebox{-0.8cm}{\includegraphics[width=2.cm]{clover}}}{\includegraphics[width=2.cm]{triangle}}
 \coleq 
-\frac{23 \pi^2}{48}\, N_2^2
\end{equation}
Indeed in the color limit implied by the conjecture, these terms coincide. This is an important partial check, because, on the one hand it does not depend on the contribution of clover diagrams, and on the other hand it confirms the advocated presence of additional terms from clover integrals implicit in the diagrams \eqref{eq:extraclover}.

\paragraph{Result for vanishing spin}
Adding the final contribution from the sextic potential diagram \eqref{eq:clover} I obtain for the un-normalized structure constant of the three-point function of chiral primary twist-one operators \eqref{eq:3ptstruc}
\begin{equation}\label{eq:final}
\frac{\tilde{\cal C}^{(2)}}{\tilde{\cal C}^{(0)}} = -\frac{\pi^2}{6} \left(N_1^2 + N_2^2 - 2\right) + \left(\frac{\pi ^2}{2} -2 \right)\left(N_1 N_2 - 1 \right)
\end{equation}
and after normalizing by the two-point functions
\begin{equation}\label{eq:finalnorm}
\frac{{\cal C}^{(2)}}{{\cal C}^{(0)}} = \frac{\pi^2}{12} \left(N_1^2 + N_2^2 - 2\right) + \left(\frac{\pi ^2}{2} -2 \right)\left(N_1 N_2 - 1 \right)
\end{equation}
The leading $N_2$ component at two loops is remarkably in agreement with the matrix model prediction \eqref{eq:prediction}.

\paragraph{Three-point functions with a spinning operator}
Finally, I comment more on the three-point functions \eqref{eq:3ptstructure}, involving an operator of spin $j$. The partial results of table \ref{tab:3pt}, obtained with the integration on insertion point method are likely to need additional corrections as is the case at vanishing spin. 
If they come only from clover integrals, then they affect the $\pi^2$ term of the structure constant exclusively, leaving the rational part uncorrected. Since this term is independent of $j$ for a large combination of diagrams, it is reasonable to believe that it stays constant for the final result as well. This would mean that the $\pi^2$ part of the structure constants of table \ref{tab:3pt} should receive corrections from clover integrals that produce the same $\pi^2$ constant as for the $j=0$ case.

Such clover corrections are more difficult to determine than for the $j=0$ case, because of the following reason.
As I mentioned before, the clover diagram itself vanishes identically for $j>0$, therefore such terms may only come from the reduction of the other diagrams. 
The additional clover integrals that in the scalar case were generated directly by numerator algebra in the integrands of the diagrams \eqref{eq:extraclover} (namely integrands which reduce to the clover form after cancelling propagators with factors in the numerator, but before performing IBP's due to the additional derivatives of the spinning operator) also vanish for $j>0$. This is due to the same mechanism that drives the clover diagram to 0, namely the particular structure of derivatives implied by the definition \eqref{eq:twist1} when applied on a bubble yields a vanishing result. Then, the only explanation for extra clover-like contributions comes from the tensor (IBP) reduction of the additional derivatives involved in the computation of operators with spin. By definition these can express certain topologies as sums of master integrals with fewer propagators. These could in principle contain the clover integral, but it vanishes identically after integrating on an insertion point.

I checked that such terms may indeed occur by considering the corresponding whole three-point function (namely without integrating on an insertion point) in momentum space and performing IBP reductions. In a simple exemplar case I indeed found that terms are generated, proportional to the clover integral, whose coefficient coincides with the corresponding tree-level three-point function at spin $j$. This is in line with my expectations, as it triggers additional terms that after taking the ratio with the tree-level structure constant are indeed independent of $j$.
This shows that there are in fact potential sources of additional terms to compensate for the desired extra $\pi^2$ terms in the structure constants involving an operator with spin.
Incidentally, when this analysis is performed on the diagrams at spin 0, it reveals no additional clover integral contribution, beyond those already found after the numerator algebra (but with no IBP reductions).
A thorough computation of this terms turns out to be quite demanding, computationally, even at low values of the spin and I have not performed it.

To summarize, I used the analysis outlined above to justify the possibility that clover-like intergals are generated that could correct the $\pi^2$ term in the structure constants with spin, in such a way that those terms are independent of the spin and coincide with the $j=0$ case, which I was able to compute more firmly.
Taking into account these additional contributions, my final estimate for the normalized structure constants is summarized in table \ref{tab:3ptj} for the $N_1 N_2$ color component
\begin{table}[h!]
\begin{tabular}{c|cccccc}
$j$ & 1 & 2 & 3 & 4 & 5 & 6 \\[2pt] \hline \\[-12pt]
$\frac{{\cal C}_{j,N_1 N_2}^{(2)}}{{\cal C}_j^{(0)}}$ & $-1+\frac{\pi^2}{2}$ & $-\frac{5}{3}+\frac{\pi^2}{2}$ & $-\frac{151}{90}+\frac{\pi^2}{2}$ & $-\frac{544}{315}+\frac{\pi^2}{2}$ & $-\frac{32561}{18900}+\frac{\pi^2}{2}$ & $-\frac{179981}{103950}+\frac{\pi^2}{2}$  \\[15pt] 
\end{tabular}
\caption{The two-loop normalized structure constants of \eqref{eq:3ptstructure}}\label{tab:3ptj}
\end{table}

and
\begin{equation}\label{eq:finalj}
\frac{{\cal C}_{j,N_1^2}^{(2)}}{{\cal C}_j^{(0)}} = \frac{\pi^2}{12}
\end{equation}
where the color components are separated according to \eqref{eq:colornot}.

\section{Conclusions}

In conclusion, I have performed the computation of three-point functions of twist-one operators in ABJM from two different angles.
I have related a color limit of the three-point function of chiral primary length-two operators to the expectation value of a supersymmetric Wilson loop and used its conjectural exact expression in terms of a matrix model to derive a two-loop result for it.
I have performed a perturbative test of this prediction by computing the three-point function at two loops in perturbation theory at weak coupling. In such a computation I have highlighted some subtleties of the method which I have used and compared my result to a previous calculation (finding a discrepancy).
In the process of computing the three-point function of chiral primary operators I have also extended the evaluation to three-point functions involving one twist-one un-protected operator with spin, thereby providing some additional perturbative data (up to spin 6). The final result for the structure constant of three protected twist-one operators is found in \eqref{eq:finalnorm} and those of three-point functions with a spinning operator are collected in table \ref{tab:3ptj} and \eqref{eq:finalj}.

An agreement can be finally established between the two-loop perturbative result and the prediction from the analysis of the matrix model describing the latitude Wilson loop.
Depending on the perspective, this test provides non-trivial support, from the matrix model \eqref{eq:matrixlat}, in favor of the result for the two-loop three-point function \eqref{eq:final} and its generalizations with a spinning operator. Or conversely it backs the validity of the matrix model proposal \eqref{eq:matrixlat}, which is still conjectural as providing the expectation value of the supersymmetric Wilson loop \eqref{eq:latitude} in ABJM, and the procedure for extracting the subset of diagrams relevant for the comparison with the three-point function. 

Despite the successful checks, given that some steps of the derivation are a bit shaky or conjectural, it would be comforting to possess an independent estimate of the three-point functions I computed.
An OPE analysis would be effective for this task, however there are no explicit computations of higher-point correlators of such operators in ABJM to the required perturbative order. Only a one-loop computation of a four-point function has been explicitly performed \cite{Bianchi:2011rn}, which only provides consistency with the vanishing of one-loop three-point functions (and with scattering amplitudes and light-like Wilson loops \cite{Chen:2011vv,Bianchi:2011dg,Henn:2010ps}).

\bibliographystyle{JHEP}

\bibliography{biblio2}

\end{document}